\documentclass[fleqn,usenatbib]{mnras}

\usepackage[T1]{fontenc}

\DeclareRobustCommand{\VAN}[3]{#2}
\let\VANthebibliography\thebibliography
\def\thebibliography{\DeclareRobustCommand{\VAN}[3]{##3}\VANthebibliography}

\newcommand{\rhm}{r_{\rm hm}}

\usepackage{graphicx}	
\usepackage{amsmath}	
\usepackage{amssymb}	
\usepackage{tabularx}
\usepackage{soul}
\usepackage{csvsimple}
\usepackage{lscape}
\usepackage{comment}

\newcommand{\msun}{{\rm M}_\odot}	

\usepackage[usenames,dvipsnames]{xcolor}
\usepackage[normalem]{ulem}
\usepackage{multirow}

\def\mgadd#1{{\textcolor{orange}{#1}}}

\newcommand{\reff}{R_{\rm eff}}	
\newcommand{\pc}{{\rm pc}}	
\newcommand{\kms}{{\rm km/s}}	
\usepackage{newtxtext,newtxmath}
\usepackage[skip=0pt]{caption}

\title[Black holes in the Hyades?]{Stellar-mass black holes in the Hyades star cluster?}

\author[S. Torniamenti et al.]{
S. Torniamenti$^{1,2,3}$\thanks{E-mail: stefano.torniamenti@unipd.it}, M. Gieles$^{4,5}$, Z. Penoyre$^{6,7}$, T. Jerabkova$^{8}$, L. Wang$^{9,10}$, F. Anders$^4$
\\
$^{1}$Physics and Astronomy Department Galileo Galilei, University of Padova, Vicolo dell'Osservatorio 3, I--35122, Padova, Italy\\
$^{2}$INFN - Padova, Via Marzolo 8, I--35131 Padova, Italy\\
$^{3}$INAF - Osservatorio Astronomico di Padova, Vicolo dell'Osservatorio 5, I-35122 Padova, Italy\\
$^{4}$ Institut de Ci\`{e}ncies del Cosmos (ICCUB), Universitat de Barcelona (IEEC-UB), Mart\'{i} i Franqu\`{e}s 1, E08028 Barcelona, Spain\\
$^{5}$ ICREA, Pg. Llu\'{i}s Companys 23, E08010 Barcelona, Spain\\
$^{6}$ Institute of Astronomy, University of Cambridge, Madingley Road, CB3 0HA, Cambridge, UK\\
$^{7}$ Leiden Observatory, Leiden University, PO Box 9513, 2300 RA, Leiden, The Netherlands\\
$^{8}$ European Southern Observatory, Karl-Schwarzschild-Stra{\ss}e 2, 85748 Garching bei München, Germany \\
$^{9}$ School of Physics and Astronomy, Sun Yat-sen University, Daxue Road, Zhuhai, 519082, China\\
$^{10}$ CSST Science Center for the Guangdong-Hong Kong-Macau Greater Bay Area, Zhuhai, 519082, China }

\date{Accepted XXX. Received YYY; in original form ZZZ}

\pubyear{2015}

\begin{document}
\label{firstpage}
\pagerange{\pageref{firstpage}--\pageref{lastpage}}
\maketitle

\begin{abstract}
Astrophysical models of  binary-black hole mergers in the Universe require a significant fraction of stellar-mass black holes (BHs) to receive  negligible natal kicks to explain the gravitational wave detections. This implies that BHs should be retained even in open clusters with low escape velocities ($\lesssim1~$km/s). 
We search for signatures of the presence of BHs in the nearest open cluster to the Sun -- the Hyades -- by comparing density profiles of direct $N$-body models to data from  {\it Gaia}. The observations are best reproduced by models with $2-3$ BHs at present. Models that never possessed BHs have an half-mass radius $\sim30\%$ smaller than the observed value, while those where the last BHs were ejected recently ($\lesssim150~$Myr ago) can still reproduce the density profile. 
In 50\% of the models hosting BHs, we find BHs with stellar companion(s). 
Their period distribution peaks at $\sim10^3$ yr, making them unlikely to be found through velocity variations. We look for potential BH companions through large \textit{Gaia} astrometric and spectroscopic errors, identifying 56 binary candidates - none of which consistent with a massive compact companion. 
Models with $2-3$ BHs have an elevated {central} velocity dispersion, but observations can not yet discriminate. 
We conclude that the present-day structure of the Hyades  requires a significant fraction of BHs to receive natal kicks smaller than the escape velocity of $\sim 3 \, \mathrm{km \, s^{-1}}$ at the time of BH formation and that the nearest BHs to the Sun are in, or near, Hyades.
\end{abstract}



\section{Introduction}
The  discovery of binary black holes (BBH) mergers with gravitational wave (GW) detectors  \citep{2021arXiv211103606T} has led to an active discussion on the origin of these systems \citep[for example,][]{2016Natur.534..512B, 2016MNRAS.458.2634M, 2016PhRvD..93h4029R,2022Natur.603..237S}. 
A popular scenario is that BBHs form dynamically in the centres of globular clusters \citep[GCs, for example, ][]{spz2001,2020PhRvD.102l3016A} and open clusters \citep[OCs, for example,][]{2019MNRAS.483.1233R, 2019MNRAS.487.2947D, 2020MNRAS.495.4268K,2021MNRAS.500.3002B,torniamenti2022}.  This scenario has gained support from the discovery of accreting BH candidates in an extragalactic GC \citep[][]{maccarone2007} and 
several Milky Way GCs \citep[][]{strader2012, chomiuk2013,millerjones2015} as well as the discovery of three detached binaries with BH candidates in the Milky Way GC NGC\,3201 \citep{giesers2018, giesers2019} and one in the 100 Myr star cluster NGC\,1850 in the Large Magellanic Cloud (\citealt{saracino2022}, but see \citealt{elbadry2022,2023MNRAS.521.3162S}).

Various studies have also pointed out that populations of stellar-mass BHs may be present in GCs, based on their large core radii \citep{2007MNRAS.379L..40M, 2008MNRAS.386...65M}; the absence of mass segregation of stars in some GCs \citep{peuten2016, 2016ApJ...833..252A, 2020ApJ...898..162W}; the central mass-to-light ratio \citep[][]{2019MNRAS.482.4713Z,2019MNRAS.488.5340B, 2019MNRAS.483.1400H,2023arXiv230301637D}; the core over half-light radius \citep{2018MNRAS.478.1844A, 2020ApJS..247...48K} and the presence of tidal tails \citep[][]{2021NatAs...5..957G}.

Recently, \citet{2021NatAs...5..957G} presented direct $N$-body models of the halo GC Palomar 5. This cluster is unusually large ($\sim20\,$pc) and is best-known for its extended tidal tails. Both these features can be reproduced by an $N$-body model that has at present $\sim20\%$ of the total mass in stellar-mass BHs. They show that the half-light radius, $\reff$, is a strong increasing function of the mass fraction in BHs ($f_{\rm BH}$). Because all models were evolved on the same orbit, this implies that the ratio of $\reff$ over the Jacobi radius is the physical parameter that is sensitive to $f_{\rm BH}$. 

At the present day, all of the searches for BH populations in star clusters focused on old ($\gtrsim10\,$Gyr) and relatively massive ($\gtrsim10^4\,\msun$)  GCs in the halo of the Milky Way, and there is thus-far no work done on searches for BHs in young OCs in the disc of the Milky Way. The reason is that most methods that have been applied to GCs are challenging to apply to OCs: for mass-to-light ratio variations, precise kinematics are required, which is hampered by orbital motions of binaries \citep{2015AJ....150...97G,rastello2020} and potential escapers \citep[][]{2000MNRAS.318..753F, 2017MNRAS.466.3937C, 2019MNRAS.487..147C} at the low velocity dispersions of OCs (few 100\,m/s). 

In the last few years, the advent of the ESA \textit{Gaia} survey (\citealt{GaiaCollaboration2016}, see \citealp{gaiadr3} for the latest release) has allowed us, for the first time, to study in detail the position and velocity space of OCs (for example, see \citealp{cantatgaudin2022} for a recent review), and to identify their members with confidence. Several hundreds of new objects have been discovered (for example, \citealp{cantatgaudin2018,cantatgaudin2018b, castroginard2018, castroginard2020, castroginard2022, sim2019, liu2019, Hunt2021, hunt2023, chi2023}), and could be distinguished from non-physical over-densities that were erroneously listed as OCs in the previous catalogues (\citealp{cantatgaudin2020}).

The possibility to reveal the full spatial extension of OCs members has made it feasible to describe in detail their radial distributions, up to their outermost regions \citep{2022A&A...659A..59T}, and to study them 
as dynamical objects interacting with their Galactic environment. 
In particular, OCs display extended halos of stars, much more extended than their cores, which are likely to host a large number of cluster members (\citealp{nilakshi2002,meingast2021}). Also, evidence of structures that trace their ongoing disruption, like tidal tails, has been found for many nearby OCs, like the Hyades (\citealp{reino2018, roser2019, lodieu2019, meingast2019, Jerabkova2021}), Blanco 1 (\citealp{zhang2020}), Praesepe (\citealp{roser2019b}), and even more distant ones like UBC 274 \citep{Piatti2020, Casamiquela2022}. 
This wealth of data provides the required information to characterize the structure of OCs in detail and, possibly, to look for the imprints given by the presence of dark components, in the same way as done for GCs. 

In this exploratory study, we aim to find constraints on the presence of BHs in the Hyades cluster, the nearest - and one of the most widely studied - OCs. We use the same approach as in the Palomar 5 study of \citet{2021NatAs...5..957G}, hence a good understanding of the behaviour of $R_{\mathrm{eff}}$ at the orbit of the Hyades is required, that is, the model clusters need to be evolved in a realistic Galactic potential. To this end, we explore the large suite of $N-$body models by \cite{wang2021}, conceived to model the impact of massive stars (that is, the BH progenitors) on the present-day structure of Hyades-like clusters. By comparing these models to the radial profiles of Hyades members with different masses from \textit{Gaia} (\citealp{evans2022}), we aim to constrain if a BH population  is required. 


The paper is organised as follows. In Sect. \ref{sec:methods}, we describe the details of the $N-$body models and our method to compare them to observations. In Sect. \ref{sec:results}, we report the results for the presence of BHs in the Hyades. In Sect. \ref{sec:discussion} we report a discussion on BH-star candidates in the cluster.
Finally, Sect. \ref{sec:conclusions} summarises our conclusions.

\section{Methods} \label{sec:methods}
\subsection{The Hyades cluster} \label{sec:observations}
The Hyades is the nearest OC to us, at a distance $d \approx 45$ pc (\citealp{perryman98}). By relying on 6D phase-space constraints, \cite{roser2011} identified $724$ stellar members moving with the bulk Hyades space velocity, with a total mass of $435 \, \mathrm{M_{\odot}}$ (\citealp{roser2011}).  The tidal radius is estimated to be $r_{\mathrm{t}} \approx 10\,  \pc$, and the resulting bound mass is $ \approx 275 \, \mathrm{M_{\odot}}$ \citep{roser2011}. Also, the cluster displays prominent tidal tails, which extend over a distance of 800 pc (\citealp{Jerabkova2021}).

The Hyades contains stars with masses approximately between 0.1 $\msun$ and 2.6 $\msun$. \cite{roser2011} found that average star mass of the cluster decreases from the center to the outward regions, as a consequence of mass segregation. Recently, \cite{evans2022} performed a detailed study of the Hyades membership and kinematics, with the aim to quantify the degree of mass segregation within the cluster.  
In particular, they applied a two-component mixture model to the \textit{Gaia} DR2 data \citep{gaiaDR2} and identified the cluster and tail members with masses $m>0.12 \, \mathrm{M_{\odot}}$ (brighter than $m_{\mathrm{G}} < 14.06$). They assigned a mass value to each observed source through a nearest-neighbour interpolation on the {\it Gaia} colour-magnitude space ($\mathrm{BP-RP}$ vs. $m_{\mathrm{G}}$). Finally, they defined two components, named “high-mass” and “low-mass” stars, based on a color threshold at $\mathrm{BP-RP}=2$, corresponding to 0.56 $\mathrm{M}_{\odot}$. The component median masses are 0.95 $\mathrm{M}_{\odot}$ and 0.32 $\mathrm{M}_{\odot}$, respectively. These values were taken as nominal masses for the two components.

Because of mass segregation, this two-component formalism has turned out to be required to adequately describe the radial cumulative mass profiles over the entire radius range and within the tidal radius \citep{evans2022}. In particular, the mass distributions of the stellar components within 10 pc are well described by a superposition of two \cite{plummer1911} models. Table \ref{tab:plummer} reports the parameters of the best-fit Plummer model (\citealp{evans2022}). 
The estimated total mass and half-mass radius of stars inside the tidal radius are $M_{\mathrm{l}}=71.9 \, \mathrm{M_{\odot}}$ and $r_{\mathrm{hm,l}}=5.7$ pc for the low-mass component, and $M_{\mathrm{h}}=170.5 \, \mathrm{M_{\odot}}$ and $r_{\mathrm{hm,h}}=4.16$ pc for the high-mass stars.

In this work, we will use the density profiles given by the best-fit Plummer models reported in Tab. \ref{tab:plummer} as observational points to compare to our $N-$body models. For this reason, hereafter we will refer to these best-fit profiles as to "observed profiles".

\begin{table}
    \centering
    \begin{tabular}{|l|r r|r r|}
    \hline 
        \multirow{2}{*}{} & \multicolumn{2}{c}{Plummer parameters} & \multicolumn{2}{c|}{Stars within 10 pc} \\ \cline{2-5}
         & $M_{\mathrm{p}}  \, \rm{(M_{\odot})} $ & $a_{\mathrm{p}} \, \rm{(pc)} $ & $M \, \rm{(M_{\odot})}$ & $r_{\rm{hm}} \, \rm{(pc)}$ \\ \hline \hline
        
         Low-mass & 117.3 & 6.21  & 71.9 & 5.67 \\ 
         High-mass & 207.5 & 3.74 & 170.5  & 4.16 \\ \hline 
    \end{tabular}
    \caption{
Left: Total mass scale ($M_{\mathrm{p}}$) and radius scale ($a_{\mathrm{p}}$) for the two components of the best-fit Plummer model, from \protect\cite{evans2022}. Right: the resulting  mass ($M$) and half-mass radius ($r_{\mathrm{hm}}$)  for the stars within 10 pc, obtained by truncating the best-fit Plummer models at  $r_{\mathrm{t}}= 10 $ pc.} \label{tab:plummer}
\end{table}

\begin{figure*}
	\includegraphics[width=\textwidth]{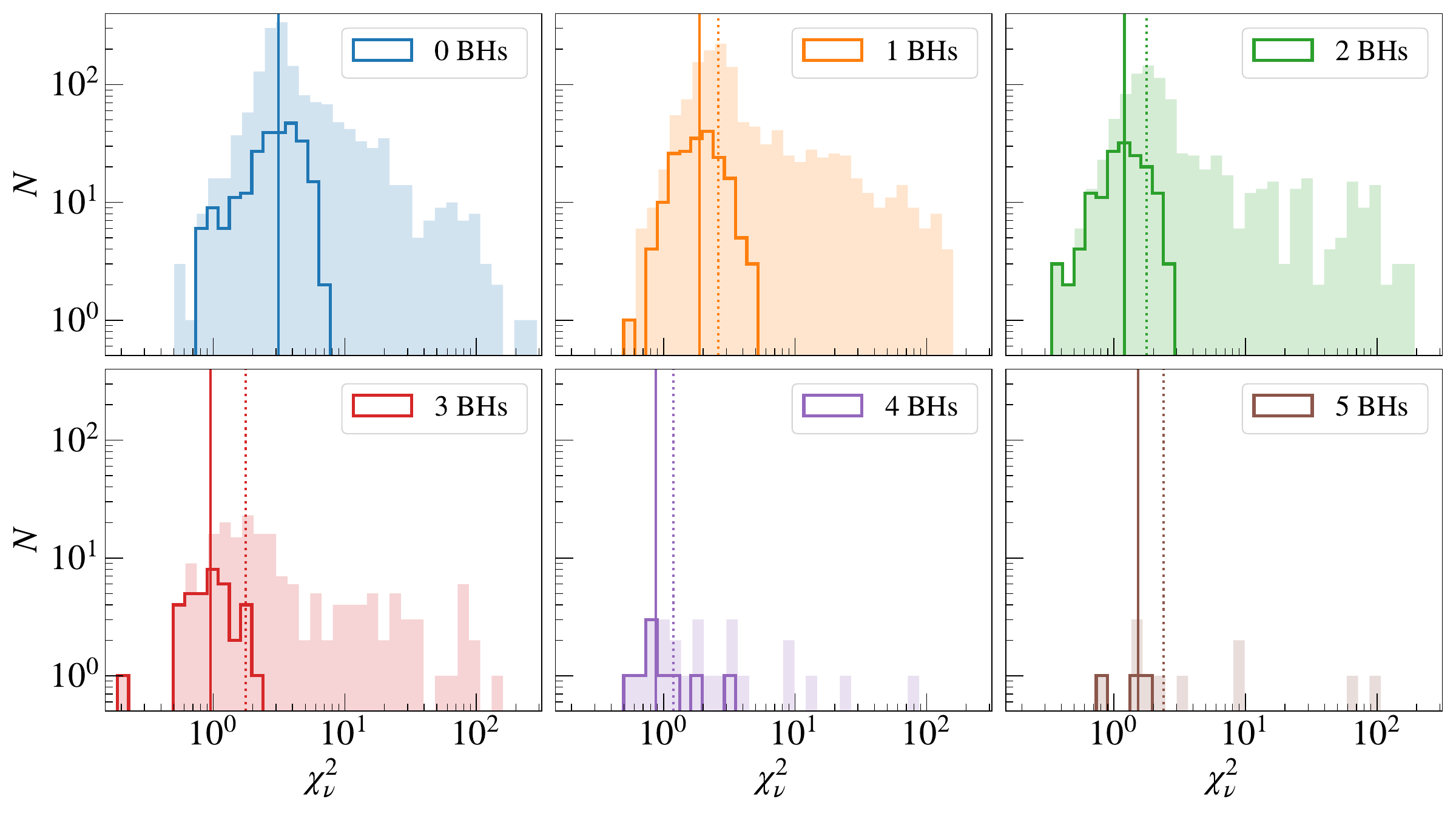}
    \caption{{Distributions of $\chi_\nu^2$ from the fits to the density profiles for star clusters with different numbers of BHs in the Hyades at the present day. The filled area include the entire distributions of star clusters, while the solid line displays the star clusters with $150 \, \mathrm{M_{\odot}}\le M_{\rm h}\le190 \, \mathrm{M_{\odot}}$. The vertical lines show the median value of the distributions when all the clusters are considered (dotted line) and when the mass cut is applied (solid line). In the models with 0 BHs, the two lines overlap.}}    \label{fig:chi_squared_Nbh}
\end{figure*}

\subsection{{\it N-}body models} \label{sec:nbody}
We use the suite of $N-$body simulations introduced in \cite{wang2021}, which aim to describe the present-day state of the Hyades cluster. 
The simulations are generated by using the $N$-body code \textsc{PeTar} \citep{Wang2020a,Wang2020b}, which can provide accurate dynamical evolution of close encounters and binaries.
The single and binary stellar evolution are included through the population synthesis codes \textsc{sse} and \textsc{bse} \citep{Hurley2000,Hurley2002,Banerjee2020}.

The ``rapid'' supernova model for the remnant formation and material fallback from \cite{Fryer2012}, along with the pulsational pair-instability supernova from \cite{Belczynski2016}, are used. 
In this prescription, if no material falls back onto the compact remnant after the launch of the supernova explosion, natal kicks are drawn from the distribution inferred from observed velocities of radio pulsars, that is a single Maxwellian with $\sigma= 265 \, \mathrm{km \, s^{-1}}$ (\citealp{hobbs2005}). For compact objects formed with fallback, kicks are lowered proportionally to the fraction of the mass of the stellar envelope that falls back ($f_{\mathrm{b}}$). In this case $v_{\mathrm{kick,fb}} = (1-f_{\mathrm{b}}) v_{\mathrm{kick}}$, 
where $v_{\mathrm{kick}}$ is the kick velocity without fallback. For the most massive BHs that form via direct collapse ($f_{\mathrm{b}}=1$) of a massive star, no natal kicks are imparted. In this formalism, the kick is a function of the fallback fraction, and not of the mass of the compact remnant. In this recipe and for the adopted metallicity of $Z=0.02$, about $45\% \, (50\%)$ of the formed BH number (mass) has $f_{\rm b}=1$, and therefore does not receive a natal kick.

The tidal force from the Galactic potential is calculated through the \textsc{galpy} code \citep{Bovy2015} with the \textsc{MWPotential2014}. This prescription includes a power-law density profile with an exponential cut-off for the bulge, a \cite{miyamoto1975} disk and a NFW profile \citep[][]{navarro1995} for the halo.

\begin{figure*}
	\includegraphics[width=\textwidth]{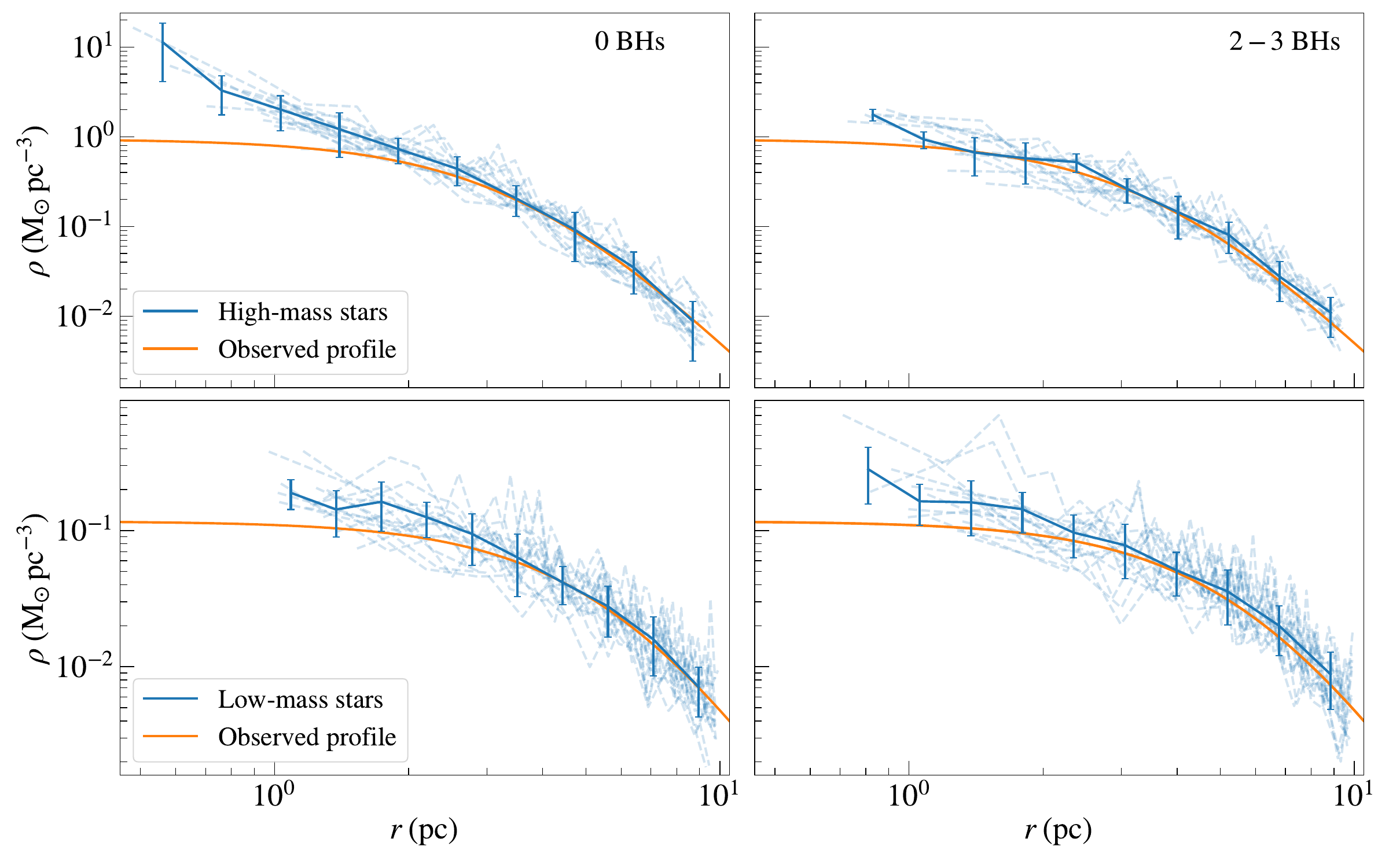}
    \caption{
    {Density profiles for high-mass stars (upper panels) and low-mass stars (lower panels), for 16 models drawn from the cases with $N_{\mathrm{BH}}=0$ (left) and $N_{\mathrm{BH}}=2-3$ (right). The blue dashed lines are the individual models. The blue solid line is the median of the distribution at selected radial distances, with the associated errors. The Plummer uncertainties are comparable to those of the $N$-body models. The orange line is the observed profile (\citealp{evans2022}). 
    }
    }
    \label{fig:density_profiles}
\end{figure*}

\subsubsection{Initial conditions}
The suite of $N-$body models consists of 4500 star clusters, initialized with a grid of different total masses $M_{0}$ and half-mass radii $r_{\rm{hm,0}}$. The initial values for $M_{0}$ are set to 800, 1000, 1200, 1400, or 1600 $\mathrm{M_{\odot}}$, while $r_{\rm{hm,0}}$ takes values 0.5, 1, or 2 pc. The initial positions and velocities are sampled from a \cite{plummer1911} sphere, truncated at the tidal radius (see below). 

The cluster initial mass function (IMF) is sampled from a \cite{kroupa2001} IMF between $0.08 - 150 \, \mathrm{M_{\odot}}$. For each couple $[M_{0},r_{\rm{h,0}}]$, \cite{wang2021} generate 300 models by randomly sampling the stellar masses with different random seeds. 
On the one hand, this allows to quantify the impact of stochastic fluctuations in the IMF sampling, which, for clusters with a limited number of particles, are generally large (for example, see \citealp{goodman1993,boekholt2015,wang2021b}). 
On the other hand, different random samplings result in different fractions of  O-type stars with $m>20 \, \mathrm{M_{\odot}}$ (the BH progenitors), which deeply affect the cluster global evolution (see \citealp{wang2021}). 

In the models considered, the mass fraction of O-type stars $f_{\rm{O}}$ ranges from to 0 to 0.34 (the expected fraction for the chosen IMF is 0.13). 
The stochasticity of the mass sampling may result in clusters with $f_{\rm{O}}=0$, meaning that they do not contain stars massive enough to form BHs at all. The percentage of clusters with $f_{\rm{O}}=0$ depends on the initial cluster mass, and varies from $6\%$ for clusters with $M_0=800 \, \msun$ to $0.7\%$ for clusters with $M_0=1600 \, \msun$. Overall, $2.4\%$ of the clusters do not host stars with $m>20 \, \mathrm{M_{\odot}}$.
No primordial binaries are included in the simulations (see the discussion in Sect.~\ref{ssec:pbh}). 
 
All the clusters are evolved for 648 Myr, the estimated age of the Hyades (\citealp{wang2021}). The initial position and velocity of the cluster are set to match the present-day coordinates in the Galaxy (see \citealp{gaia2018HR,Jerabkova2021}). For this purpose, the centre of the cluster is first integrated backwards for 648 Myr in the \textsc{MWPotential2014} potential by means of the time-symmetric integrator in \textsc{galpy}. The final coordinates are then set as initial values for the cluster position and velocity \citep{wang2021}.
The resulting initial tidal radius is (see also Fig. 5 in \citealp{wang2022}): 
\begin{equation}
    r_{\mathrm{t,0}} \approx 12 \left[\frac{M_{\mathrm{0}}}{1000 \, \msun}\right]^{1/3} \; \mathrm{pc,}
\end{equation}
while the tidal filling factor, defined as $r_{\mathrm{hm,0}} \, / \, r_{\mathrm{t,0}}$, spans from 0.03 to 0.18. Stars that initially lie outside the tidal radius are removed from the cluster. 

\subsection{Comparing models to observations} \label{sec:comparison}

We build the model density profiles from the final snapshots of the $N-$body simulations. First, we center the cluster to the density center, 
calculated as the square of density weighted average of the positions (\citealp{casertano1985,2003gnbs.book.....A}).
Then, we build the profiles for low-mass and high-mass stars within $r_{\rm{t}}$, separately. To be consistent with the observed profiles (see Sect. \ref{sec:observations}), we define all the stars below $0.56 \, \mathrm{M_{\odot}}$ as low-mass stars, and all the luminous main-sequence and post-main sequence
stars above this threshold as high-mass stars. Also, because we want to compare to observable radial distributions, we only include the visible components of the cluster (main sequence and giant stars), without considering white dwarfs, neutron stars and BHs. We divide the stellar cluster into radial shells containing the same number of stars. Due to the relatively low number of stars, we consider $N_{\rm bin}=10$ stars per shell.

\begin{table*}
  \centering
    \begin{tabular}{c|ccccccccc} 
    \hline
    $N_{\mathrm{BH}}$  &  $M_{\rm{vis}}$ (M$_\odot$) & $M_{\rm{h}}$ (M$_\odot$)  &  $M_{\rm{tot}}$ (M$_\odot$) & $f_{\rm{BH}}$ & $f_{\rm{O}}$ & $M_{0}$ (M$_\odot$) & $r_{h,0}$ (pc) & $P_{\mathrm{cut}}$  \\ 
          \hline\hline \\
    0 BHs & $233.9^{+21.4}_{-22.1}$ & $170.5^{+12.3}_{-15.1}$  & $254.0^{+24.4}_{-24.1}$ & $0$  & $0.09_{- 0.05 }^{+0.06}$ & $1016.1_{-16.1}^{+194.5}$ & $0.98_{-0.48}^{+0.99}$ & 13.8 \\ \\
    1 BHs & $242.5^{+21.0}_{-21.9}$ & $170.5^{+15.6}_{-10.7}$  & $274.1^{+22.5}_{-25.0} $ & $0.04^{+0.02}_{-0.01}$ & $0.12_{- 0.06}^{+ 0.06}$ & $1201.4_{-200.6}^{+200.3}$ & $0.99_{-0.49}^{+0.99}$ & 13.6 \\ \\
    2 BHs & $241.2^{+21.8}_{-22.1}$ & $168.1^{+14.5}_{-11.1}$  & $280.2^{+22.9}_{-25.4}$ & $0.07^{+0.02}_{-0.02}$ & $0.15_{-0.05}^{+0.05}$  & $1401.4_{-200.6}^{+200.3}$  & $1.00_{-0.50}^{+0.99}$ & 14.2 \\ \\
    3 BHs & $242.7^{+27.6}_{-26.2}$ & $173.0^{+10.9}_{-18.0}$  & $289.6^{+30.8}_{-28.4}$ & $0.09^{+0.02}_{-0.01}$ & $0.15_{-0.04}^{+0.05}$  &  $1400.5_{-197.3}^{+195.5}$  & $1.96_{-1.27}^{+0.03}$ & 16.8 \\ \\
    4 BHs & $249.3^{+14.2}_{-29.4}$ & $167.1^{+14.1}_{-7.2}$  & $294.5^{+23.7}_{-22.4}$ & $0.11^{+0.02}_{-0.01}$ & $0.17_{-0.04}^{+0.03}$ & $1400.5_{-0.2}^{+195.3}$  & $1.97_{-0.71}^{+0.04}$ & 27.2 \\ \\
    5 BHs & $216.7^{+25.5}_{-8.5}$ & $155.6^{+6.0}_{-3.4}$  & $281.4^{+18.8}_{-14.2}$ & $0.16^{+0.01}_{-0.02}$ & $0.18_{-0.05}^{+0.02}$ & $1598.5_{-270.2}^{+0.3}$ & $1.97_{-0.02}^{+0.00}$ & 27.2  \\ \\
    \hline
    \end{tabular} 
  \caption{Properties of the Hyades models with $150 \, \mathrm{M_{\odot}}\le M_{\rm h}\le190 \, \mathrm{M_{\odot}}$, for different numbers of BHs in the Hyades at the present day ($N_{\mathrm{BH}}$, column 1): total mass in visible stars (column 2), total mass in high-mass stars (column 3), total mass (column 4), BH mass fraction (column 5), initial mass fraction in O-type stars (column 6), initial total mass (column 7), initial half-mass radius (column 8). The last column reports the percentage of models that evolve into clusters within the mass cut, for the selected $N_{\mathrm{BH}}$. The reported values are the medians of the distributions, while the subscripts and superscripts are the difference from the 16\% and 84\% percentiles, respectively.
  }\label{tab:results}
\end{table*}%

To assess how well the models reproduce the observed profiles, we refer to a 
$\chi^2$ comparison, where we define the reduced $\chi^2$, $\chi_\nu^2$ (with an expected value near 1), as:
\begin{equation}
    \chi^2_{\nu} = \frac{1}{\nu} \sum_i \frac{(\rho_{{\rm obs},i} - \rho_{{\rm mod},i})^2}{\delta \rho^2_i},
\end{equation}
where $\nu$ is the number of degrees of freedom, which depends on the number of density points obtained with the binning procedure. 
The quantities $\rho_{{\rm obs},i}$ and $\rho_{{\rm mod},i}$ are the density in the $i^{\mathrm{th}}$ bin for the observed and model profile, respectively. 
The error $\delta \rho^2_i$ is given by the sum of the model and the observed bin uncertainties. For  both observed and $N-$body profiles, we determine the  uncertainty as the Poisson error:
\begin{equation} \label{eq_uncertainty}
    \delta \rho = \frac{\bar{m}}{4/3 \, \pi \, \left(r^3_{f}-r^3_{i}\right)} \sqrt{N_{\rm bin}},
\end{equation}
where $\bar{m}$ is the mean mass of the bin stars, and $r^3_{i}$ and $r^3_{f}$ are the bin upper and lower limit. For the $N-$body models, the bin lower (upper) limit is set as the position of the innermost (outermost) star, and $\bar{m}$ is the mean stellar mass in each bin. 
For the observed profiles, we consider the same bin boundaries as the $N-$body models, and set $\bar{m}$ to the nominal mass of the component under consideration. 
Then, we estimate analytically from the \cite{plummer1911} distribution the number of stars between $r_{i}$ and $r_{f}$ and the corresponding uncertainty.

Our comparison is performed by considering the high-mass density profile only. This choice is motivated by the fact that the observed mass function in Fig. 2 of \cite{evans2022} displays a depletion below 0.2 $\msun$, which may hint at possible sample incompleteness. We thus focus only on the high-mass range to obtain a more reliable result. Also, high-mass stars, being more segregated, represent better tracers of the innermost regions of the cluster, where BHs are expected to reside, and thus provide more information about the possible presence of a dark component. We emphasize that this is intended as a formal analysis with the objective of determining whether a model is able to give a reasonable description of the observed cluster profile. 

In order to filter out the simulations that present little agreement with the observations, we consider only the models with a final high-mass bound mass within $\pm 20 \, \rm{M_{\odot}}$ from the observed value of $M_{\mathrm{h}} = 170.5 \, \rm{M_{\odot}}$ (see Tab. \ref{tab:plummer}). Among the simulated models, 636 clusters (14\% of all the $N$-body models) lie within this mass range.

\section{Results} \label{sec:results}
As the cluster tends towards a state of energy equipartition, the most massive objects progressively segregate toward its innermost regions, while dynamical encounters push low-mass stars further and further away (\citealp{spitzer1987}). BHs, being more massive than any of the 
stars, tend to concentrate at the cluster centre, quenching the segregation of massive stars. As a consequence, their presence in a given star cluster is expected to affect the radial mass distribution of the cluster’ stellar population (\citealp{fleck2006, hurley2007, peuten2016}; \mgadd{\citealt{2016ApJ...833..252A, 2020ApJ...898..162W}}) 

In the star cluster sample under consideration, the number of BHs within 10 pc, $N_{\mathrm{BH}}$, ranges from 0 to 5. 
Star clusters with $N_{\mathrm{BH}}=0$ can result from the ejection of all the BHs, because of supernovae kicks (50\% of the cases) and/or as the result of 
dynamical interactions. 
As for supernovae kicks, since our $N-$body models have initial escape velocities $v_{\mathrm{esc}} \lesssim 6 \, \mathrm{km \, s^{-1}}$, which decrease to $v_{\mathrm{esc}} \lesssim 3 \, \mathrm{km \, s^{-1}}$ at 24 Myr, only BHs formed with  kicks lower than $3\, \mathrm{km \, s^{-1}}$ can be retained (see also \citealp{pavlik2018}). Also, as mentioned earlier, the IMF may not contain stars massive enough to form BHs (12\% of the models within the mass cut that end up with 0 BHs, see Sect. \ref{sec:nbody}). 

In the following, we will assess if $N_{\rm{BH}}\leq 5$ BHs can produce quantifiable imprints on the radial distributions of stars.

\subsection{\texorpdfstring{$\chi_{\nu}^{2} $}{} distributions}

Fig. \ref{fig:chi_squared_Nbh} shows the distributions of 
$\chi_{\nu}^{2}$ for different $N_{\mathrm{BH}}$. 
If we apply the mass cut introduced in Sect. \ref{sec:comparison}, we automatically select most of the models with $\chi_{\nu}^{2}$ closer to the expected value near 1, and remove those that are highly inconsistent with the observed profiles. 
The result of our comparison improves with increasing the number of BHs up to $N_{\mathrm{BH}}= 4$, which however applies to only $1\%$ of the cases. If we focus on the cases with a large number of good fits ($N_{\mathrm{BH}} \leq 3$), the median value of the reduced chi-squared distributions decrease from $\chi_{\nu}^{2}  \approx 3$ to $\chi_{\nu}^{2}  \approx 1$ for $N_{\mathrm{BH}}$ increasing from 0 to 3.

When only models within the mass cut are considered, they have $N_{\mathrm{BH}} \leq 3$ in 98\% of the cases. 
This is mainly because star clusters that contain a high initial mass fraction in O-type stars (which evolve into BHs) are easily dissolved by the strong stellar winds (\citealp{wang2021}), and result in present-day cluster masses far below the observed one. If the initial mass fraction in O-type stars is more than twice as high as that expected from a \cite{kroupa2001} IMF, our models cannot produce clusters in the selected mass range. 


Table \ref{tab:results} reports the final relevant masses and mass fractions of the $N-$body models,  for different values of $N_{\mathrm{BH}}$.
In all the cases, the total mass in high-mass stars is $\approx 170 \, \rm{M_{\odot}}$, as a consequence of the chosen criterion for filtering out models with little agreement with the observed cluster. The total visible mass, $M_{\rm{vis}} \approx 240 \, \mathrm{\msun}$, does not show any dependence on $N_{\mathrm{BH}}$, with the only exception of the sample with 5 BHs. For the latter case, as mentioned earlier, the initial larger mass fraction of O-stars brings about a more efficient mass loss across the tidal boundary, and results in lower cluster masses. In contrast, the total mass $M_{\rm{tot}}$ increases with $N_{\rm{BH}}$: the mass in BHs spans from $\approx 10 \, \mathrm{\msun}$ ($f_{\rm{BH}}=0.04$) when $N_{\rm{BH}}=1$, to $\approx 45 \, \mathrm{\msun}$ for the case with 5 BHs ($f_{\rm{BH}}=0.16$). 

\subsection{Two-component radial distributions } \label{sec:radial_distribution}

To highlight the difference between models with and without BHs, we randomly draw 16 models from simulations (within the mass cut) with 0 BHs and from a sample obtained by combining the sets with 2 and 3 BHs. For each distribution, we evaluated the median values for selected bins and the spread, as $1.4 \times \mathrm{MAD} \, \left(\sqrt{N_{\mathrm{bin}}}\right)^{-1}$, where $\mathrm{MAD}$ is the median absolute deviation.

Fig. \ref{fig:density_profiles} displays the density profiles of the high-mass (top) and low-mass (bottom) stars of these samples, compared to the observed profiles (see Sect. \ref{sec:observations}). The density profiles of the $N-$body models with BHs are mostly consistent with the observed distributions. High-mass stars in clusters with $N_{\mathrm{BH}}=0$ display a more concentrated distribution reminiscent of the cusped surface brightness profiles of core collapsed GCs \citep{1986ApJ...305L..61D}. The models with BHs have cored profiles, which  \citet{2004ApJ...608L..25M} attributed to the action of a BH population. Although in our models there are only 2 or 3 BHs, it has been noticed already by \citet{hurley2007} that a single BBH is enough to prevent the stellar core from collapsing. It is worth noting that the Plummer models that were fit to the observations are cored and would therefore not be able to reproduce a cusp in the observed profile. But from inspecting the cumulative mass profile in Fig. 3 of \citet{evans2022} we see that the observed profile follows the cored Plummer model very well, with hints of a slightly faster increase in the inner 1 pc of the high-mass components, compatible with what we see in the top-right panel of Fig.~\ref{fig:density_profiles}. 

\begin{figure}
	\includegraphics[width=\columnwidth]{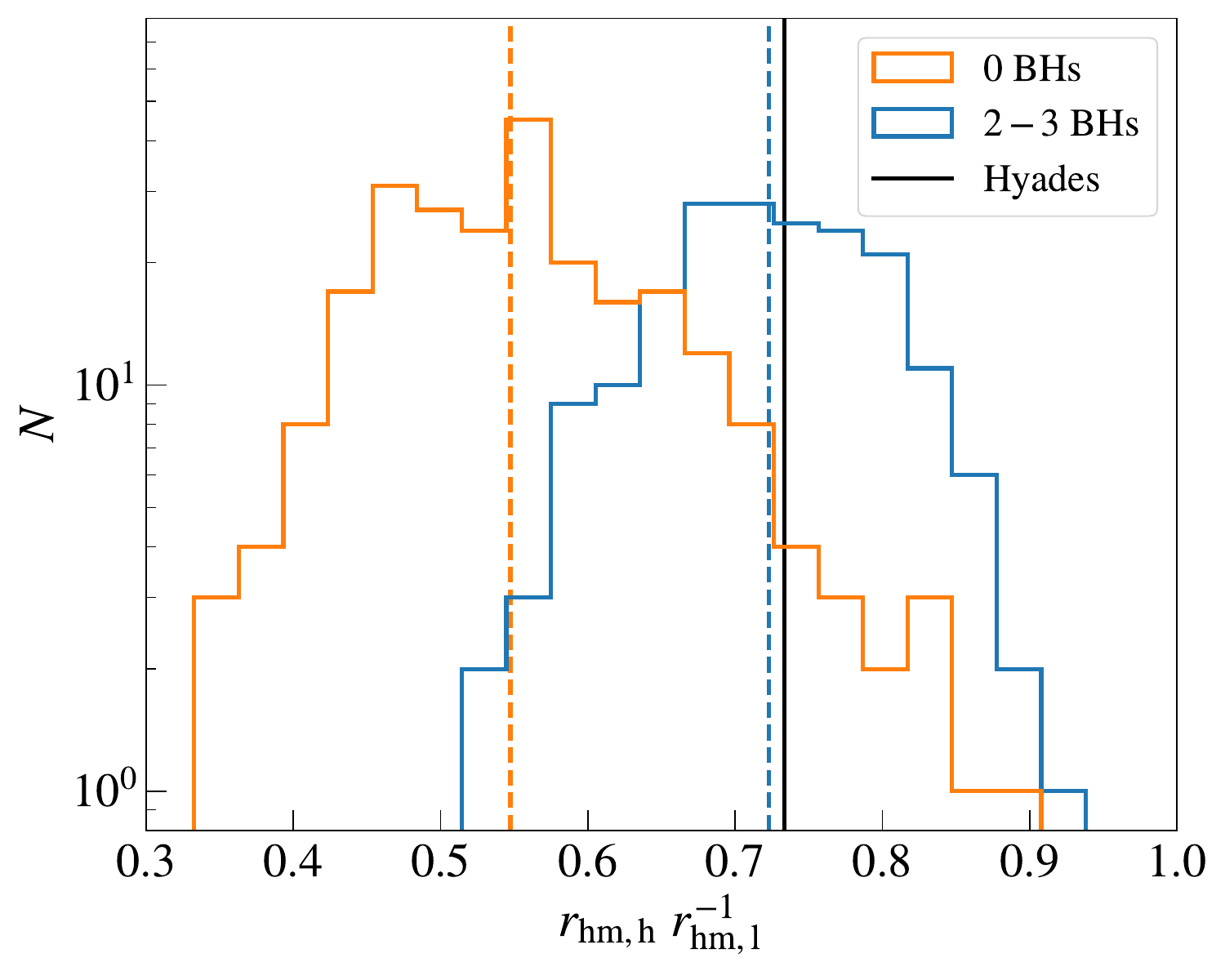}
    \caption{Ratio of the half-mass radius of the high-mass stars ($r_{\rm{hm,h}}$) to that of low-mass stars ($r_{\rm{hm,l}}$), for star clusters with $N_{\mathrm{BH}}=0$ (orange) and $N_{\mathrm{BH}}=2-3$ (blue). The dashed vertical lines represent the medians of the distributions, and the vertical black line displays the observed value for the Hyades (\citealp{evans2022}).  }
    \label{fig:rhl_rhh}
\end{figure}

The density profile of low-mass stars is also well described by models with BHs, although they were not included in the fitting procedure. This component presents central densities lower than high-mass stars of about an order of magnitude, as a consequence of mass segregation within the cluster.
A better description of the relative concentration of stars with different masses (and thus of the degree of mass segregation) is given by the ratio of their half-mass radii (for example, see \citealp{vesperini2013,vesperini2018,devita2016,torniamenti2019}). 
Figure \ref{fig:rhl_rhh} displays the ratio of the half-mass radius\footnote{In this study, the half-mass radii are calculated from the distributions of the stars within $r_{\rm{t}}$, and do not refer to the half-mass radii of the whole Plummer model.} of high-mass to that of low-mass stars, for all the models with 0 BHs and with 2$-$3 BHs. For the latter case, BHs produce less centrally concentrated distributions 
of visible stars, and trigger a lower degree of mass segregation. Also, models with BHs yield a much better agreement with the observed value.

\begin{figure*}
	\includegraphics[width=\textwidth]{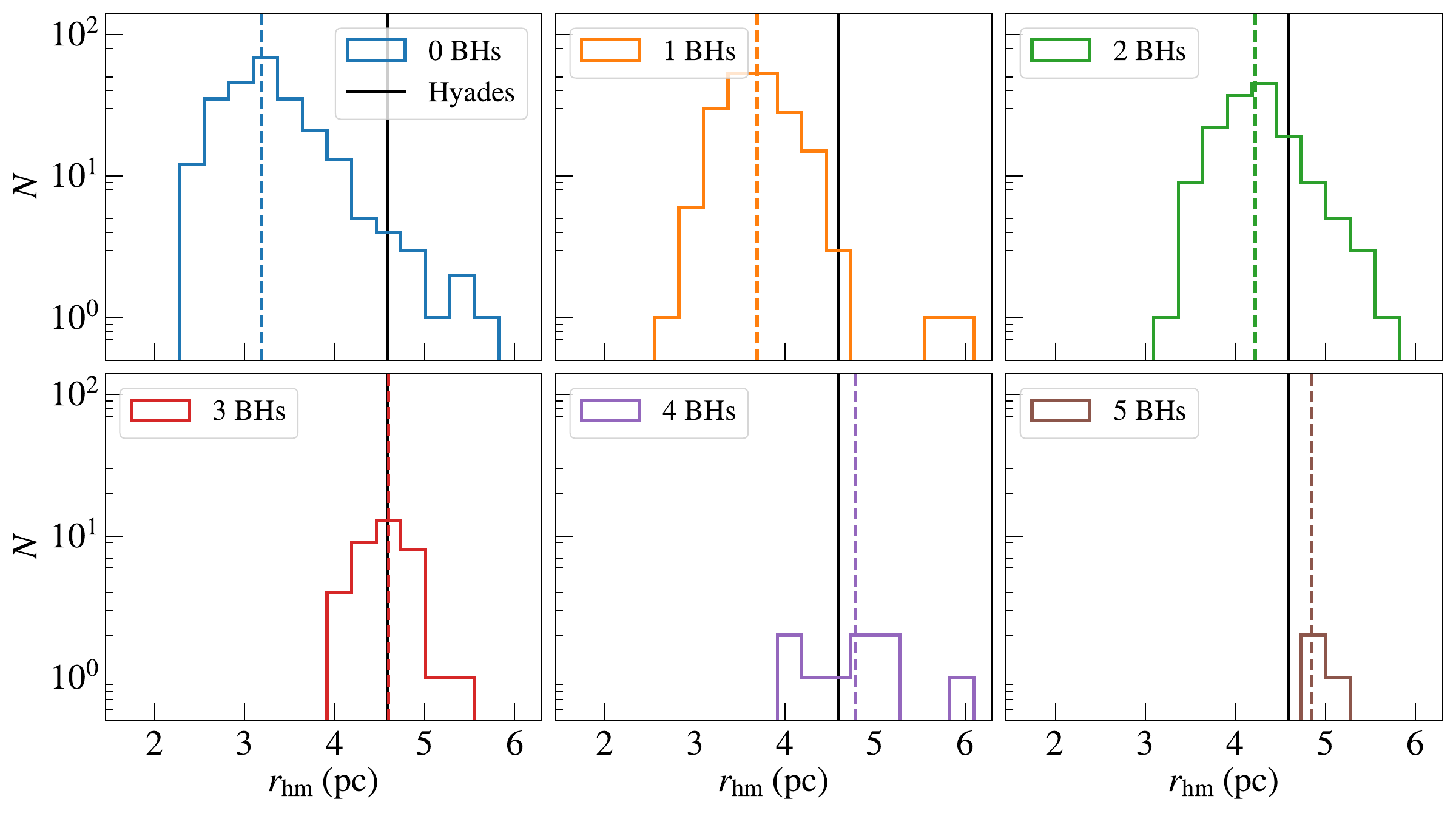}
    \caption{
    Distributions of half-mass radii of visible stars for $N-$body models with different $N_{\mathrm{BH}}$. The dashed vertical lines represent the medians of the distributions, and the vertical black line displays the observed value for the Hyades (\citealp{evans2022}). } 
    \label{fig:rhm_Nbh}
\end{figure*}

\subsection{Half-mass radii} \label{sec:half_mass}

Figure \ref{fig:rhm_Nbh} shows the impact of BHs on $r_{\rm hm}$, defined as the half-mass radius of all the visible stars. The distributions shift towards higher values for increasing numbers of BHs, which is because $r_{\rm hm}$ is larger, but also because of the quenching of mass segregation of the visible components. Our models suggest that 3 BHs can produce a $\sim 40 \%$ increase in the expected value of $r_{\rm hm}$. As a further hint on the presence of a BH component, the observed value almost coincides with the expected value for $N_{\rm{BH}}=3$.

The $r_{\rm hm}$ distribution of the $N_{\rm{BH}}=0$ sample is mostly inconsistent with the observed value of the Hyades cluster. Unlike the other cases, this distribution shows a more asymmetric shape, with a peak at $\rhm\simeq3\,$pc, and a tail which extends towards larger values. 
We investigated if this tail may come from clusters that have recently ejected all their BHs, and have still memory of them. 
Figure \ref{fig:rhm_time} shows the the distribution of the half-mass radii for the cases without BHs at the present day. We distinguished between different ranges of $t_{\rm{BH}}$, defined as the time at which the last BH was present within the cluster. 
The stellar clusters that have never hosted BHs, because they are ejected by the supernova kick or because there are no massive stars to produce them (see Sect. \ref{sec:results}), constitute the bulk of the distribution. 
These models end up to be too small with respect to the Hyades, and thus are not consistent with the observations, regardless of their $M_0$ and $r_{\rm hm,0}$ (see also the discussion in Sect. \ref{ssec:ics}).
From Fig.~\ref{fig:rhm_time} we also see that the $N-$body models where all the BHs were ejected in the first 500 Myr show the same $r_{\rm hm}$ distribution as those that have never hosted BHs. For these clusters, the successive dynamical evolution has erased the previous imprints of BHs on the observable structure, because the most massive stars had enough time to segregate to the center after the ejection of the last BH. 

Finally, star clusters where BHs were present 
in the last $\sim150\,$ Myr, but are absent at present, preserved some memory of the ejected BH population, and display larger $r_{\rm hm}$, in some cases consistent with the observed value. Since the present-day relaxation time (\citealp{spitzer1987}) for our $N-$body models is $t_{\mathrm{rlx}}\approx 45$ Myr, we find that the only models that have ejected their last BH less than $3 \, t_{\mathrm{rlx}}$ ago 
can have radii similar to models with BHs. 

BHs that were ejected from the Hyades in the last 150 Myr display a median distance $\sim 60$ pc from the cluster ($\sim 80$ pc from the Sun). Only in two cases, the dynamical recoil ejected the BH to a present-day distance $> 1$ kpc, while in all the other cases the BH is found closer than $200$ pc from the cluster center.

\subsection{High-mass stars parameter space} \label{sec:high_mass}

As explained in Sect. \ref{sec:half_mass}, the presence of even $2-3$ BHs has a measurable impact on the observable structure of such small-mass clusters. High-mass stars are most affected by the presence of BHs, because they are prevented from completely segregating to the cluster core. 
In Fig. \ref{fig:rhh_Mhh} we show how the number of BHs within the cluster relates to the total mass in high-mass stars ($M_{\rm{h}}$) and to their half-mass radius ($r_{\rm{hm,h}}$). 
In this case, we consider all the simulated models, without any restriction on the high-mass total mass, and we show how the average number of BHs in the $N-$body models varies in the $M_{\rm{h}}-r_{\rm{hm,h}}$ space. 

The total mass in high-mass stars can be as high as 400 $\mathrm{M_{\odot}}$, while the half-mass radius takes values from 1 to 8 pc. The most diluted clusters feature the lowest mass, because they are closer to being disrupted by the Galactic tidal field. In contrast, models with higher $M_{\rm{h}}$ are characterized by the fewest BHs, because of the absence of massive progenitors, which enhance the cluster mass loss. As explained in Sect. \ref{sec:radial_distribution}, $r_{\rm{hm,h}}$ grows for increasing number of BHs at the cluster center. In the Hyades mass range, the expected value of $r_{\mathrm{hm,h}}$ when $N_{\mathrm{BH}}=3$ is larger by almost $\sim 60 \%$ with respect to the case with 0 BHs.
The observed values (\citealp{evans2022}) lie in a region of the parameter space between 2$-$3 BHs, a further corroboration of the previous results of Sect. \ref{sec:results}. Finally, higher numbers of BHs are disfavoured by our models, because they predict an even lower degree of mass segregation for high-mass stars.

\subsection{Velocity dispersion profiles}

We quantified the impact of central BHs on the velocity dispersion profile. To this purpose, we compared the profiles obtained from the samples of 16 models with $N_{\mathrm{BH}}=0$ and with $N_{\mathrm{BH}}=2-3$ introduced in Sect. \ref{sec:radial_distribution}. Figure \ref{fig:sigma} displays the resulting velocity dispersion profiles, calculated as the mean of the dispersions of the three velocity components. The presence of 2$-$3 BHs produces a non-negligible increase of 40\% in the inner 1 pc. 

The rise in dispersion is reminiscent of the velocity cusp that forms around a single massive object \citep{1976ApJ...209..214B}. Such a cusp  develops within the sphere of influence of a central  mass, which can be defined as $G M_{\bullet}/\sigma^2$, with $M_\bullet$ the mass of the central object and $\sigma$ the stellar dispersion. For $M_\bullet=20\,\msun$ and $\sigma =0.3\,\kms$ we find that this radius is $\sim1\,\pc$, roughly matching the radius within which the dispersion is elevated. Although a BBH of $20\,\msun$ constitutes $\sim10\%$ of the total cluster mass, the mass with respect to the individual stellar masses is much smaller (factor of $20$) compared to the case of an intermediate-mass BH in a GC (factor of $10^4$) or a super-massive BH in a nuclear cluster (factor of $10^6$). As a result, a BBH in Hyades makes larger excursions from the centre due to Brownian motions. From eq.~90 in \citet{2001ApJ...556..245M} we see that the wandering radius of a BBH of $20\,\msun$ in Hyades is $\sim0.15\,\pc$. Although this is smaller than the sphere of influence, it is still a significant fraction of this radius. We therefore conclude that the elevated dispersion is due to the combined effect of stars bound to the BBH, stars being accelerated by interaction with the BBH \citep{2005MNRAS.364.1315M} and the Brownian motion of its centre of mass.

The average increase of the velocity dispersion profile in the innermost parsec for models containing BHs indicates the potential for further validation through observations. 
Studies estimating the velocity dispersion of the Hyades provide central values as low as $0.3$ km/s \citep{Madsen03, Makarov2000}, and upper limits of $0.5$ km/s \citep{Douglas19} and 0.8 km/s \citep{roser2011}. 
The \textit{Gaia} data membership selection is often a trade-off between completeness and contamination and, especially for low-mass evolved star clusters, it requires a special case. 
For example, the data sets from \cite{Jerabkova2021} or \cite{roser2019}, who aimed to detect the extended tidal tails of the Hyades, may not be the ideal for the construction of the velocity dispersion profile. 

Since a detailed comparison between theoretical and observed velocity dispersion profiles requires a dedicated membership selection and a thorough understanding of the involved uncertainties, we will leave it to a follow-up focused study.
Moreover, the $N-$body models by \cite{wang2021} do not consider primordial binary stars (see discussion in Sect.~\ref{ssec:pbh}), which might affect the calculated velocity dispersion.

\begin{figure}
	\includegraphics[width=\columnwidth]{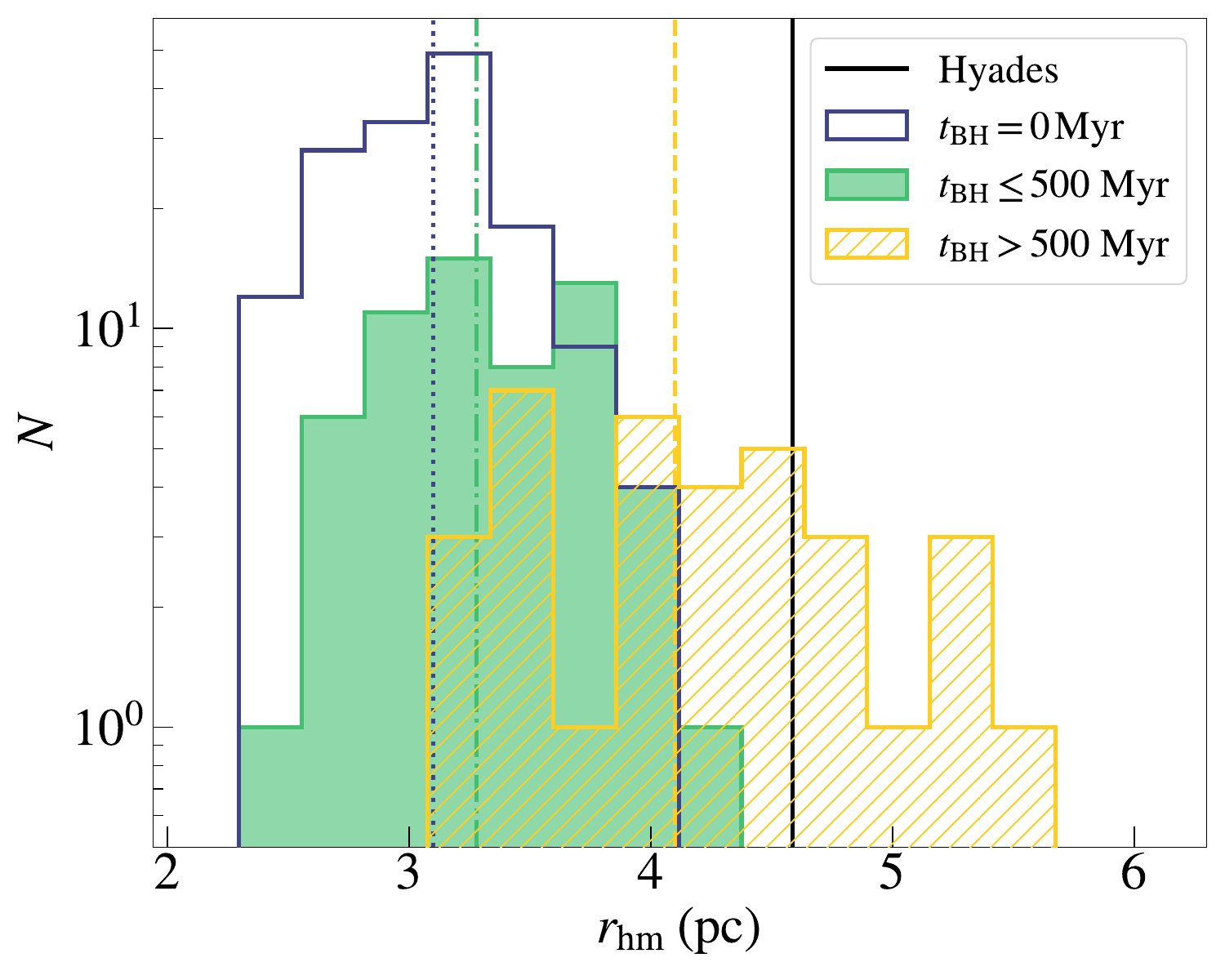}
    \caption{
    Distributions of $r_{\rm{hm}}$ for star clusters with no BHs. We distinguish between $N-$body models where BHs have never been present, because they were ejected by their natal kicks or there were not stars massive enough (purple, vertical dotted line), star clusters were BHs were ejected before 500 Myr (green filled area, vertical dash-dotted line), and star clusters were BHs were ejected after 500 Myr (yellow hatched area, vertical dashed line). The black line displays the value derived from observations (\citealp{evans2022}).
    }
    \label{fig:rhm_time}
\end{figure}

\begin{figure}
	\includegraphics[width=\columnwidth]{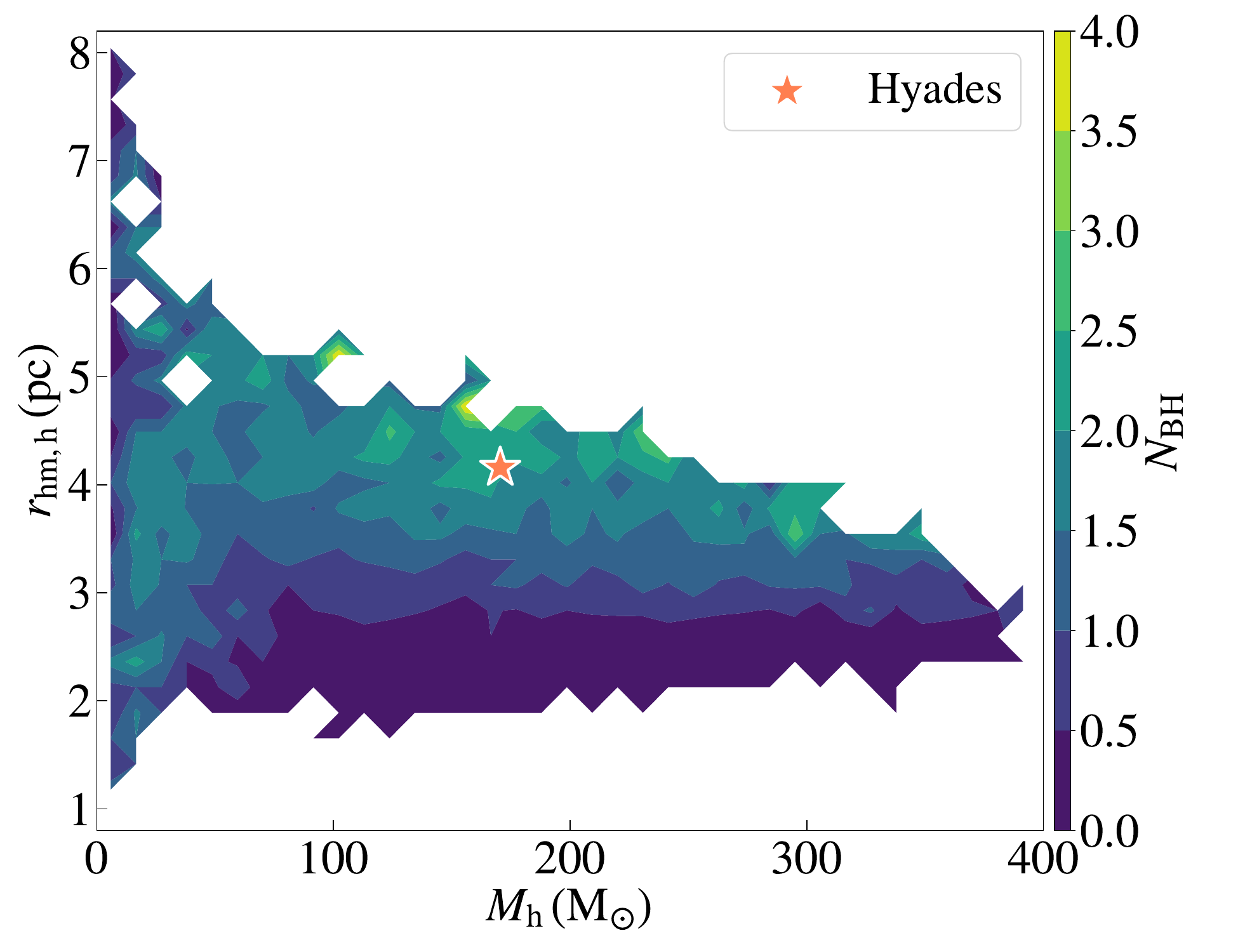}
    \caption{Contour plot of the total mass ($M_{\rm{h}}$) and the half-mass radius ($r_{\rm{hm,h}}$) of the high-mass stars. The colormap encodes the local mean number of BHs in that region of the parameter space. The orange star displays the values derived from observations (\protect\citealp{evans2022}).
    }
    \label{fig:rhh_Mhh}
\end{figure}

\begin{figure*}
	\includegraphics[width=\textwidth]{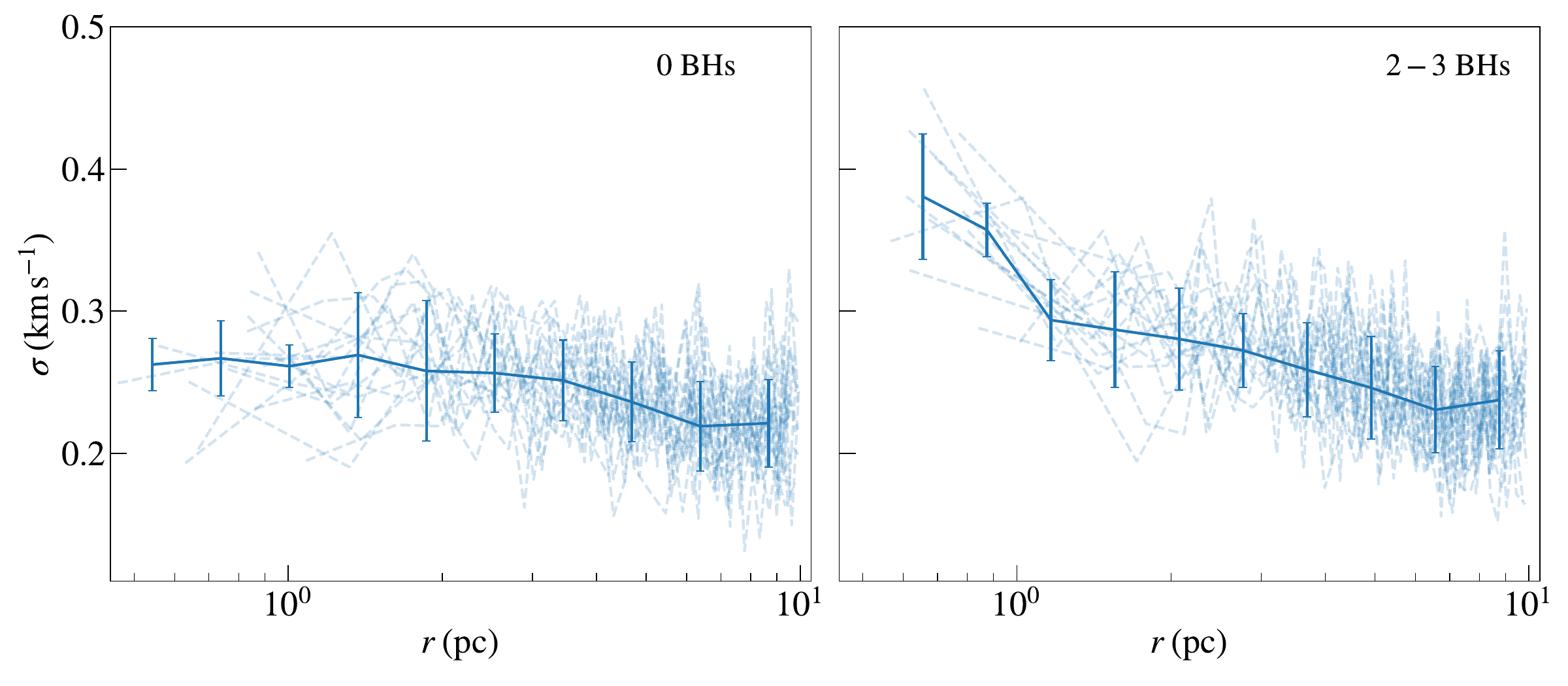}
    \caption{
    One-dimensional velocity dispersion profiles for 16 models drawn from the cases with $N_{\mathrm{BH}}=0$ (left) and $N_{\mathrm{BH}}=2-3$ (right). The blue dashed lines are the single models. The blue solid line is the median of the distribution at selected radial distances, with the associated errors.
    }
    \label{fig:sigma}
\end{figure*}

\subsection{Dynamical mass estimation}
Based on the stellar mass and the velocity dispersion, \citet{oh2020} concluded that the Hyades is super-virial and therefore disrupting on an internal crossing timescale. 
The measured velocity dispersion within the cluster is commonly used to calculate the dynamical mass of the cluster, as: 
\begin{equation} \label{eq:mdyn}
M_{\mathrm{dyn}} \simeq \frac{10 \, \langle \sigma_{\rm 1D}^2\rangle \, R_{\rm eff}} {G}.   \end{equation}
We apply this to our $N$-body models and compare it to the actual total mass. To be consistent with observations, we defined $\sigma_{\mathrm{1D}}$ as the line-of-sight velocity dispersion of high-mass stars and the effective radius $R_{\mathrm{eff}}$ as the radius containing half the number of high-mass stars. We find a systematic bias of $M_{\mathrm{dyn}}$  overestimating the total mass of the cluster typically by a  factor of $\sim 1.5$ for $N_{\mathrm{BH}}=0$ and a factor of $\sim 2$ for $N_{\mathrm{BH}}>0$. This is  due to the presence of energetically unbound stars that are still associated with the cluster, the so-called potential escapers \citep{2000MNRAS.318..753F}, whose fraction increases as the fraction of the initial stars remaining within the cluster decreases \citep{2001MNRAS.325.1323B}.

In our $N-$body models, the clusters in the Hyades mass range (within the selected mass cut) typically retain a fraction $\sim 0.2$ of the initial stars. For these models, the percentage of potential escapers increases from $\lesssim 5\%$ in the initial conditions to $\sim 40\%$ at the present day. The fraction of potential escapers is similar to that found in  \cite{2017MNRAS.466.3937C} for models initialized with a \cite{kroupa2001} IMF (between 0.1 and 1 $\msun$) that evolve in a Galactic potential similar to the cusp of a Navarro-Frenk-White \citep{navarro1995} potential, the same adopted for the dark matter halo in the \textsc{MWPotential2014} (see Sect. \ref{sec:nbody}).
If we do not include the potential escapers in the calculation of the dynamical mass (eq. \ref{eq:mdyn}), we find values that are consistent with the actual total mass of the cluster.
We therefore conclude that the high dispersion of Hyades is not because it is dissolving on a crossing time, but because it contains potential escapers and BHs. 

\subsection{Angular momentum alignment with BBH}
The presence of a central BBH may also affect the angular momentum of surrounding stars. In particular, three-body interactions between the central BBH and the surrounding stars can lead to a direct angular momentum transfer. As a consequence, the interacting stars are dragged into corotation, and display angular momentum alignment with the central BBH  \citep{2005MNRAS.364.1315M}. This scenario works for BBHs with massive components ($> 50 \, \msun$), which are able to affect the angular momentum distribution for a relatively high fraction of stars \citep{2005MNRAS.364.1315M}.
We tested this scenario for BBHs with components of lower masses, by considering our models of the Hyades with a central BBH. In this case, stars show isotropic distribution with respect to the central BBH, independently of the distance from the cluster center. Thus, no signature of angular momentum alignment is found.

\begin{figure*}
    \includegraphics[width=\textwidth]{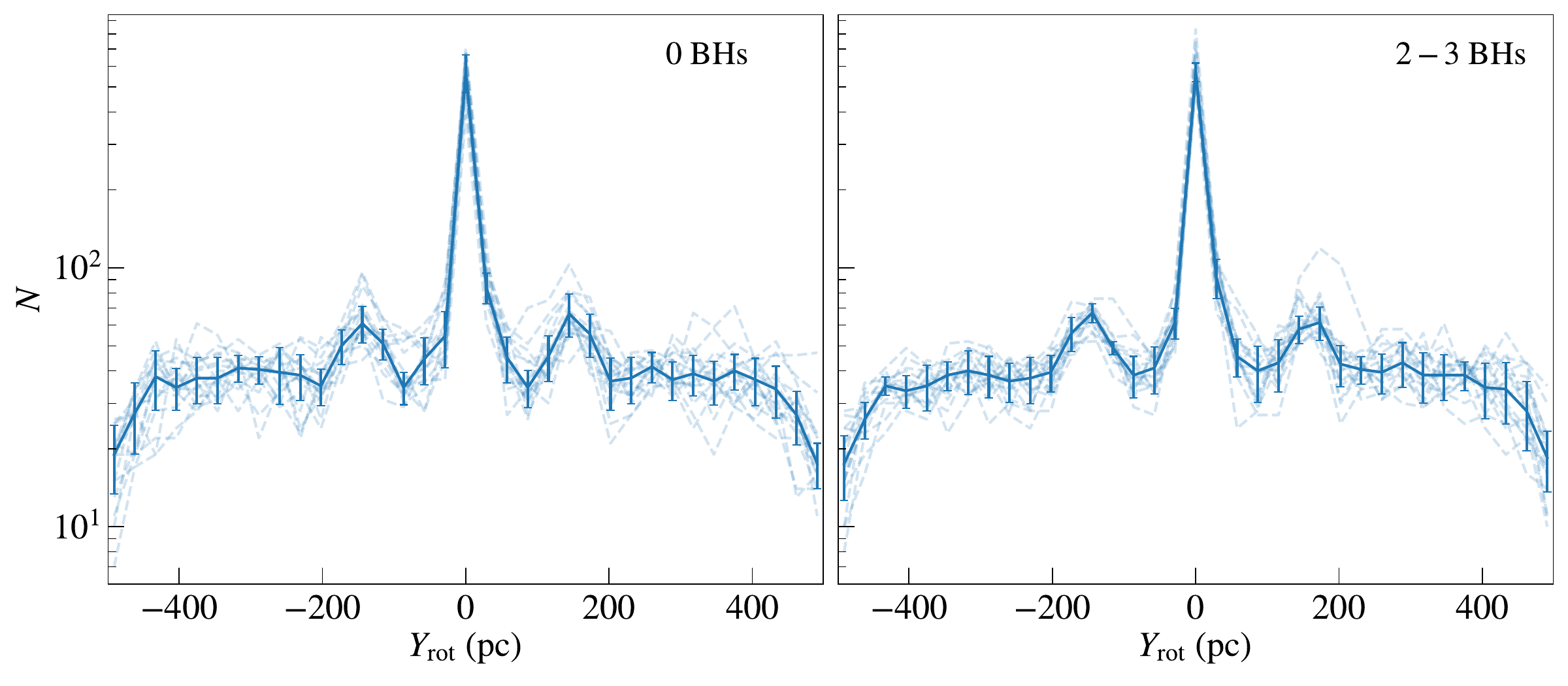}
    \caption{Tidal tail profiles for 16 models drawn from the cases with $N_{\mathrm{BH}}=0$ (left) and $N_{\mathrm{BH}}=2-3$ (right). The $Y$ Galactic coordinate is rotated, so that the $V_{\mathrm{Y}}$ component is aligned with the tail. 
    The profiles are obtained from the $N-$body models by considering all the visible stars with magnitude $m_{\mathrm{G}}<18$. 
    }
    \label{fig:tail_structure_comp}
\end{figure*}

\begin{figure*}
    \includegraphics[width=1.02\textwidth]{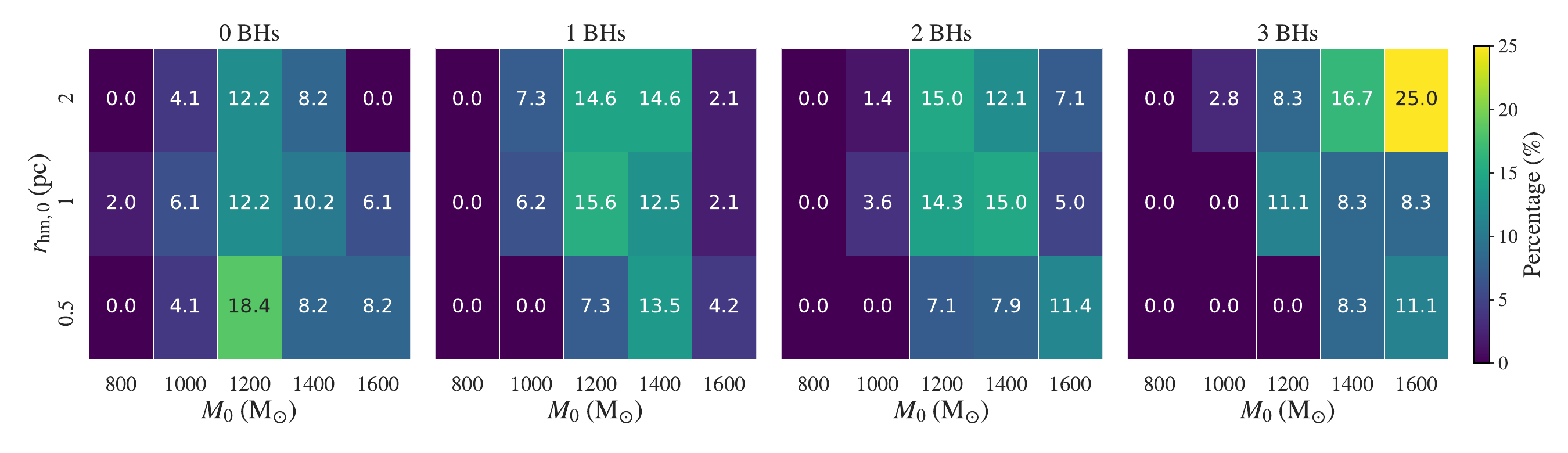}
    \caption{Percentage distributions of models that match the observations as a function of $M_{\mathrm{0}}$ and $r_{\mathrm{hm,0}}$, for different numbers of BHs in the Hyades at the present day. Here, we define the models that match the observations as those that lie within the selected mass cut (see Sect. \ref{sec:comparison}) and whose half-mass radius does not differ more than 20\% from the observed value. 
    }
    \label{fig:heatmap_ics}
\end{figure*}

\subsection{Tidal tails}
{The relaxation process  increases the kinetic energy of stars to velocities higher than the cluster escape velocity, unbinding their orbits into the Galactic field. When this mechanism becomes effective, stellar clusters preferentially lose stars through their Lagrange points \citep{kupper2008}, leading to the formation of two so-called tidal tails. The members of tidal tails typically exhibit a symmetrical S-shaped distribution as they drift away from the cluster, with over-densities corresponding to the places where escaping stars slow down in their epicyclic motion \citep{kupper2010,kupper2012}.}

Until few years ago, tidal tails had mainly been observed in GCs (\citealp[for example, see][]{odenkirchen2003} for the case of Palomar 5), which are more massive, older, and often further from the Galactic plane than OCs. Thanks to the \textit{Gaia} survey, we have now the possibility to unveil such large-scale (up to kpc) structures near OCs dissolving into the Galactic stellar field \citep[for example, ][]{roser2019, meingast2019}. 
Since the {\it Gaia} survey only provides radial velocity values for bright stars \citep{Cropper2018}, the search for tidal tail members mostly relies on projected parameters, like the proper motions, which have complex shapes. In this sense, mock observations from $N-$body models are generally adopted as a reference to recover genuine tail members, and to distinguish them from stellar contaminants \citep[for example, see][]{Jerabkova2021}.

Here, we focus on the impact of the present-day number of BHs  on the tidal tail structure. As reported in Tab. \ref{tab:results}, models with a larger number of BHs generally result from the evolution of more massive clusters ($M_0$ is $\sim 10 \%$ larger), because of the more efficient mass loss. This may produce a quantifiable impact on the number and density profile of the predicted tails. 

Figure \ref{fig:tail_structure_comp} shows the number density profiles of the tidal tails from the 16 models with 0 BHs and with $2-3$ BHs  introduced in Sect. \ref{sec:radial_distribution}. The median profiles and the associated uncertainties are built in the same way as for the density profiles.
To reduce the projection effects due to spatial alignment and emphasize the tail structure along the direction of the tail itself, we display the number of stars as a function of the $Y$ Galactic coordinate, rotated so that the $V_{\mathrm{Y}}$ component is aligned with the tail. 
Also, to obtain a sample that mimics \textit{Gaia} completeness, we consider only stars with magnitude $m_{\mathrm{G}}<18$ mag. 
The profiles of models with and without BHs are almost indistinguishable, hinting at a tiny impact from the BH content. This appears in contradiction with the fact that the initial masses of the models with BHs are 50\% higher than the models without BHs (see Tab. \ref{tab:results}), while their present-day masses are similar. 
However, $f_{\mathrm{O}}$ is also larger for clusters that retain BHs, and this leads to an enhanced mass-loss from winds in the first $\sim50$ Myr (see Fig. 5 and 7 in \citealp{wang2021}). This results in models with $N_{\mathrm{BH}}=2-3$ having a number of stars in the tails that is only $\sim 10\%$ (about 200 stars) larger than those without BHs. The recent mass-loss rates of the two sets of models is comparable. The position of the epicyclic over-densities is not affected by the number of BHs (see also Fig. 8 of \citealp{wang2021}). 

This results means that the tidal tails of clusters as low-mass as Hyades can not be used to identify BH-rich progenitors, as was suggested from the modelling of the more massive cluster Pal 5 \citep{2021NatAs...5..957G}. Future work should show whether tails of more massive OCs are sensitive to the (larger) BH content of the cluster. Also, future studies might specifically target the epicyclic over-densities in more detail and establish their phase-space properties for mode models to provide large statistical grounds. While the current observational data are not sufficient to provide such information, this will likely change with the future {\it Gaia} data releases and the complementary spectroscopic surveys SDSS-V \citep{Almeida2023}, 4MOST (\citealp{4most}) and WEAVE (\citealp{weave}).

\section{Discussion and observational tests} \label{sec:discussion}
\subsection{Dependence of the results on the initial parameters}
\label{ssec:ics}
As shown in Fig. \ref{fig:rhl_rhh} and \ref{fig:rhm_Nbh}, models with $2-3$ BHs are favored to match the observed radial distributions of the Hyades.
However, we can not use the final distributions as posteriors since the initial sampling was done on a rigid grid with fixed number of models at each grid point.
In this section, we thus explore how the choice of the initial parameters can affect our results, and if different initial values of $M_{\mathrm{0}}$ and $r_{\mathrm{hm,0}}$ would lead to a different conclusion concerning the consistency of models with 0 BHs with observations.

Figure \ref{fig:heatmap_ics} shows the percentage distributions of the models that match the observations, as a function of $M_{\mathrm{0}}$ and $r_{\mathrm{hm,0}}$, and for different values of $N_{\mathrm{BH}}$. We define such models as those that lie within the selected mass cut (see Sect. \ref{sec:comparison}) and whose half-mass radius does not differ more than 20\% from the observed value. 
For the considered $N_{\mathrm{BH}}$, we evaluate the percentage of clusters that originate from each $M_{\mathrm{0}}-r_{\mathrm{hm,0}}$ combination. 
Independently on $N_{\mathrm{BH}}$, models with $M_{\mathrm{0}}< 1000 \, \msun$ can hardly produce Hyades-like clusters. This is also evident from Fig. 6 of \cite{wang2022}, which indicates that more massive clusters are needed to reproduce the observed properties. 

Most of the models with $N_{\mathrm{BH}} <3$ lie well within the initial mass range, with lower percentages at the low- and the high-mass end. 
No clear dependence on the initial radius is found. In contrast, star clusters with 3 BHs mainly result from $M_{\mathrm{0}}$ and $r_{\mathrm{hm,0}}$ at the upper boundary of the parameter distributions. This is mainly due to the larger number of massive progenitors, which enhance the cluster mass loss, as already pointed out in Sect. \ref{sec:high_mass}. At the same time, models with larger radii retain more BHs (fewer dynamical interactions) and they therefore need to be more massive.

Our analysis suggests that more massive and extended initial conditions may produce Hyades-like clusters. However, these clusters are likely to host $N_{\mathrm{BH}} \geq 3$. Thus, a more extensive exploration of the initial parameter space is expected to strengthen the conclusion that a fraction of BHs needs to be retained within the cluster to match the observed properties of the Hyades. Furthermore, Fig. \ref{fig:rhm_time} indicates that models with no retained BHs end up too small, independently on their initial radius. Therefore, clusters with larger initial radii and no retained BHs are expected to shrink and lose mass at  a constant density \citep{1965AnAp...28..992H}, as also found for the case of Palomar 5 (see \citealp{2021NatAs...5..957G}).
As a consequence, there is no hint that, by extending the range of initial conditions, we will find different conclusions on the consistency of models with no BHs with observations.

\subsection{Possible effect of primordial binaries}
\label{ssec:pbh}
The $N-$body models considered for this work do not contain primordial binaries, but observations find that young star clusters have high binaries fractions, especially among massive stars  \citep{2012Sci...337..444S,2017ApJS..230...15M}. 
Here we discuss the possible effect of primordial binaries on the structure of clusters and, in particular, whether there may be a degeneracy with the effect of BHs. 

\cite{wang2022} investigated the impact of different mass-dependent primordial binary fractions on the dynamical evolution of star clusters with $N$-body simulations. Their results show that massive primordial binaries (component masses $> 5 \, \msun$)
dominate over low-mass binaries and that in the presence of massive binaries the evolution of the core and half-mass radius is 
insensitive to the binary fraction among low-mass stars \citep[see figure 5 in][]{wang2022}. Models with 100\% 
binaries have a $\sim10\%$ larger half-mass radius than models without binaries. This difference is less than the difference we find between clusters with and without BHs.

However, the model clusters of \citet{wang2022} are more massive ($N\sim10^5$), so they all contain some BHs. \citet{hurley2007} presents $N$-body models of clusters without 
BHs and with modest binary fractions (5\% and 10\%). The BH natal kicks are larger in his model and BH retention is therefore rare. He finds that 
the binary fraction does not affect the evolution of the core and half-mass radius. 
\citet{2011MNRAS.410.2698G} find from Monte Carlo models of 47 Tucanae that the evolution of the half-mass radius is not affected by primordial binaries. \citet{hurley2007} showed that when two BHs are retained, the effect of the BBH that inevitably forms on the observed core and half-mass radius is far larger than the primordial binaries. In particular, his Fig. 6 shows that the model with a BBH has a central surface density that is a factor of $\sim4$ lower than models with binaries and without BBH.
Given the modest binary fraction of Hyades \citep[$\sim20\%$,][]{2016A&A...585A...7K,evans2022,2023AJ....165..108B}, we therefore conclude that it is unlikely that 
primordial binaries have the same effect on the density profile as BHs. However, it would be interesting to verify this.

In conclusion, we recognize that the presence of primordial binaries play a crucial role on the long-term evolution of a cluster like the Hyades. However, a detailed characterization of the primordial binary impact on the cluster present-day structure, as well as a complete disentanglement of their observational signatures from those left by BHs, requires a more in-depth study. For this reason, we will explore it in a future work.

\subsection{BH companions}
Three-body interactions within a stellar cluster strongly favour the formation of binary systems, mainly composed of the most massive objects (\citealp{heggie1975}). As a consequence, BHs tend to form binaries preferentially with other BHs, and when in binaries with a lower-mass stellar companion, they rapidly exchange the companion for another BH (\citealp{hills1980}). 
In general, the result is a growing BBH population in the cluster core (\citealp{spz2001}). 
In OCs, however, given the limited number of BHs by the initial low number of massive stars, a non-negligible fraction of BH-star binary systems may form and survive. 

Binary stars in dynamically-active clusters are expected to display semi-major axis distributions that depend on the cluster properties. Soft binaries (with binding energy lower than the average cluster kinetic energy) are easily disrupted by any strong encounter with another passing star or binary \citep{heggie1975}. The upper limit for the semi-major axis is thus given by the hard-soft boundary of the cluster:
\begin{equation}
a_{\rm{max}} = \frac{G m_{\rm{1}} m_{\rm{2}}}{2 \, \langle m  \sigma^{2} \rangle},
\end{equation}
where $m_{\rm{1,2}}$ are the masses of the binary components, and $E_{\rm b}=\langle m\sigma^2\rangle$ is the hard-soft boundary (\citealp{heggie1975}).
For an OC with $\sigma \approx 0.5 \, \rm{km \, s^{-1}}$, the upper limit for a binary composed of a black-hole ($m_{\rm{1}} = 10 \, \rm{\msun}$) and a star ($m_{2} = 1 \, \rm{\msun}$) is of the order of $a_{\rm{max}}\sim 10^{-1}$ pc.

\begin{figure*}
\includegraphics[width=\textwidth]{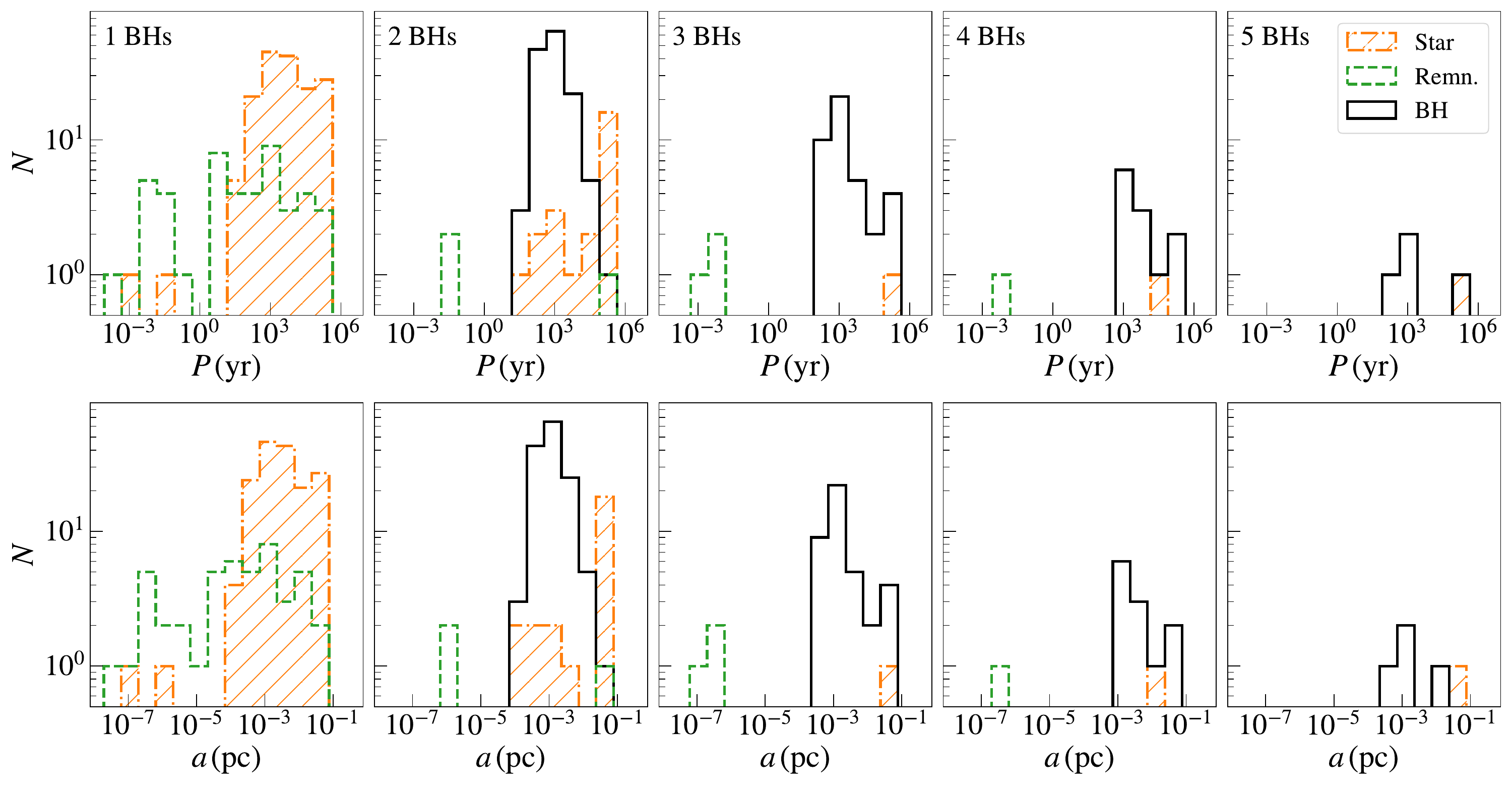}
\caption{Distributions of periods (upper panels) and semi-major axes (lower panels) of the binary and triple systems hosting BHs, for $N-$body models with different $N_{\mathrm{BH}}$. We distinguish between different types of BH companions: stars (orange dash-dot line, hatched area), white dwarfs or neutron stars (green dashed line), and BHs (black).
    }
    \label{fig:aP}
\end{figure*}

\begin{table}
  \centering
    \begin{tabular}{c|ccc} 
    \hline
    $N_{\mathrm{BH}}$  &  $f_{\rm{BH-Star}}$ &  $f_{\rm{BH-Remn.}}$ &    $f_{\rm{BH-BH}}$ \\ 
          \hline\hline 
    1 BHs & 0.78 & 0.22 & 0.0 \\
    2 BHs & 0.15 & 0.02 & 0.83 \\
    3 BHs & 0.02 & 0.07 & 0.91 \\
    4 BHs & 0.07 & 0.07 & 0.86 \\
    5 BHs & 0.2 & 0.0 & 0.8 \\
    \hline
    \end{tabular}  
\caption{Fractions of binary systems hosting BHs, for different $N_{\mathrm{BH}}$ (column 1). We distinguish between different types of BH companions: stars (column 2), white dwarfs or neutron stars (column 3), and BHs (column 4).}
\label{tab:BHcompanions}
\end{table}%

\begin{figure*}
\includegraphics[width=0.910\textwidth]{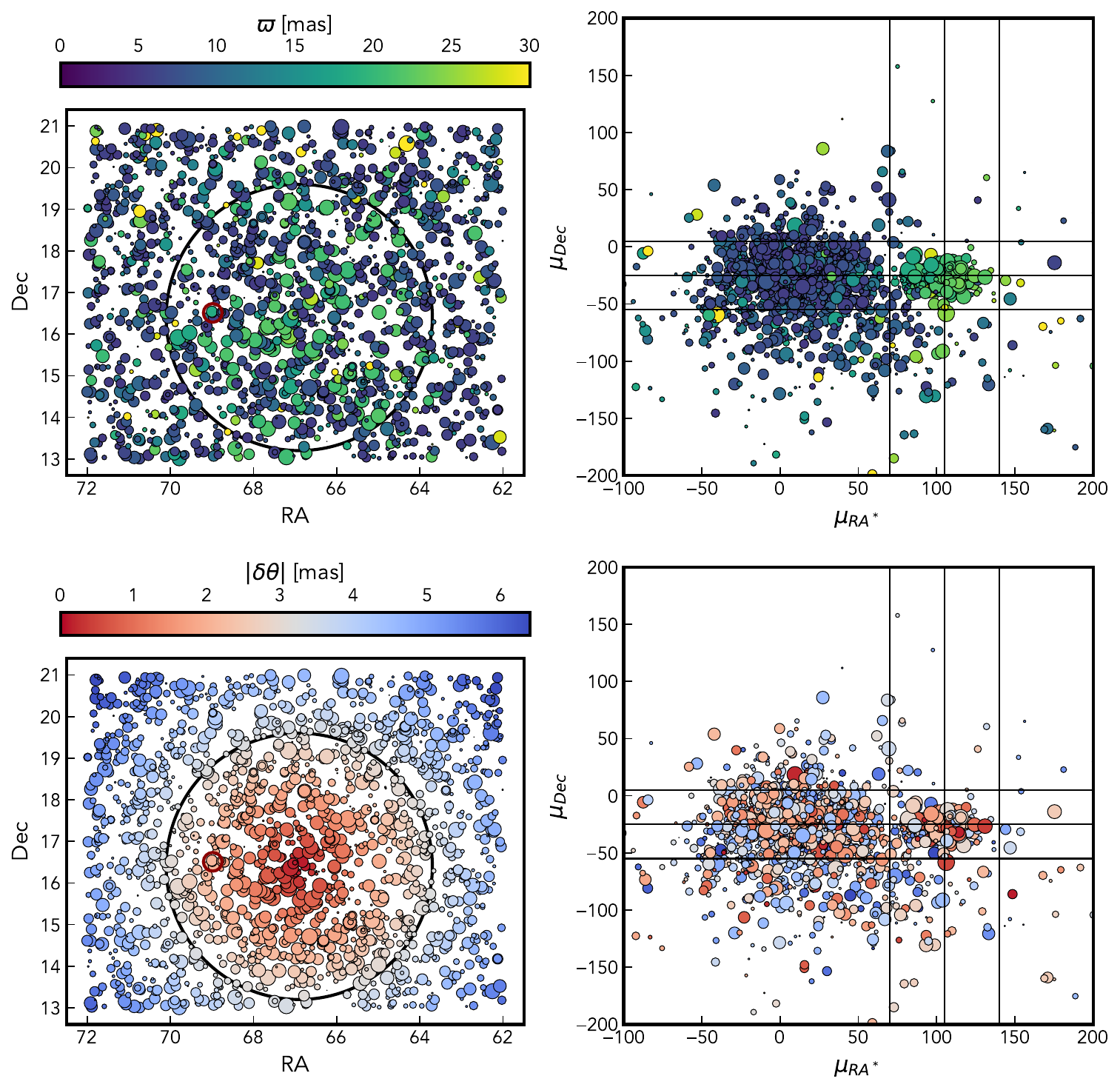}
    \caption{Position on sky (left) and proper motion (right) of sources in the field of the Hyades (with $\varpi>5$ mas). We show the parallax (top row) and angular offset from the center of the cluster (bottom row). Aldebaran, a foreground star too bright for {\it Gaia}, is shown as a red open circle. The size of each point is set by their apparent magnitude and only sources with $m_G<15$ are shown (see Fig. \protect\ref{fig:hyades_colour} for reference). We show an angular offset of 3.2 mas (black circle, left) and lines denoting $\mu_{RA^*}=105 \pm 35$ mas yr$^{-1}$ from this (black vertical lines, right) and $\mu_{Dec}=-25 \pm 30$ mas yr$^{-1}$ (black horizontal lines, right).
    }
    \label{fig:hyades_field}
\end{figure*}

When a hard binary is formed, it becomes further tightly bound through dynamical encounters with other cluster members \citep{heggie1975,goodman1984,kulkarni1993,sigurdsson1993}. Each encounter causes the binary to recoil, until the binary becomes so tight that the recoil is energetic enough to kick it out from the cluster. For this, the lower limit $a_{\rm{min}}$ can be assumed to be the semi-major axis at which the binary that produces a recoil equal to the escape velocity $v_{\rm{esc}}$. 
Following \cite{2016ApJ...831..187A}:
\begin{equation}
a_{\rm{min}} = 0.2 \frac{Gm_1m_2}{v^2_{\rm{esc}}}\frac{m_3^2}{m_{12}^2m_{123}},
\end{equation}\label{eq:amin}where $m_3 = \langle m\rangle$, $m_{12}=m_1+m_2$, and  $m_{123}=m_1+m_2+m_3$. 
For an open cluster with $v_{\rm{esc}} \approx 0.5 \, \rm{km \, s^{-1}}$, $m_1=m_2=10\,\msun$ and $m_3=0.5\,\msun$, we obtain $a_{\rm{min}}\sim 10^{-5}$ pc (2 AU). For a BH-star binary system ($m_2 = 1 \, \msun$), $a_{\mathrm{min}} \sim 10^{-4}$ pc.

BHs in our $N-$body models, as expected, show a tendency to dynamically couple with other objects, and form binary and triple systems. When $N_{\mathrm{BH}}>0$, only 6\% of the BHs are not bound in binary or multiple systems. Even in models where only 1 BH is present, the single BH tends to form binaries with (mainly) stars or other remnants (white dwarfs of neutron stars).
Figure \ref{fig:aP} shows the distribution of semi-major axes and periods for binaries and triple systems of clusters with $N_{\rm{BH}}$ ranging from 1 to 4. Independently of $N_{\rm{BH}}$, most of the binaries display semi-major axes from $10^{-5}$ pc to $10^{-1}$ pc, consistently with our approximate calculation. When more than 1 BH is present, dynamical interactions tend to favour the formation of BBHs. As reported in Tab. \ref{tab:BHcompanions}, the fraction of BBHs represents by far the largest fraction of binary systems hosting BHs if more than 1 BHs is present.

\subsection{Binary candidates in the Hyades}
In this section we present a search for possible massive companions to main sequence stars in the Hyades. We identify binary candidates by searching for members with enhanced \textit{Gaia} astrometric and spectroscopic errors (following \citealt{Penoyre2020, Belokurov2020}, and \citealt{Andrew22}).

\subsubsection{Selecting cluster members}


We start with all {\it Gaia} DR3 sources with $\varpi>5$ mas, RA between 62 and 72 degrees, Dec between 13 and 21 and $RUWE$, which stands for renormalized unit-weight error, greater than 0 (effectively enforcing a reasonable 5-parameter astrometric solution) - giving 5640 sources as shown in Fig. \ref{fig:hyades_field}. We also apply an apparent G-band magnitude cut of $m_{\rm G}<15$ above which the astrometric accuracy of {\it Gaia} starts to degrade rapidly due to Poisson noise. Analysis beyond this magnitude is eminently possible, but for such a nearby population of stars this cut excludes a minority of the cluster (even more so the likely binary systems, as binary fraction increases with mass) and means that {\it Gaia} should have a near constant ($\sim$0.2 mas, \citealt{Lindegren21}) precision per observation and thus allows uncomplicated comparison of sources.

To select cluster members we use the position, proper motion, and parallax to construct an (unnormalized) simple membership probability:
\begin{equation} \label{eq:membership_probability}
p_{\rm member} = e^{-\sum_x\left(\frac{x-x_0}{\sigma_x'}\right)^2}
\end{equation}
where
\begin{equation}
\sigma_x'^2 = \sigma_x^2 + \sigma_{AEN}^2 + \sigma_{x_0}^2
\end{equation}
with $x$ denoting each of the parameters of RA, Dec, $\mu_{RA^*} (= \mu_{RA}\cos(Dec))$, $\mu_{Dec}$ and $\varpi$. $\sigma_x$ is the reported uncertainty on each parameter in the {\it Gaia} catalog and $\sigma_{AEN}$ is the \texttt{astrometric\_excess\_noise} (AEN) of the fit. $x_0$ and $\sigma_{x_0}$ are the assumed values and spread of values expected for the cluster as listed in Tab. \ref{tab:hyades_params}. The inclusion of the AEN ensures that potentially interesting binaries, which may have a significantly larger spread in their observed values and thus fall outside of the expected variance of the cluster, are not selected against.

\begin{figure}
\centering
\includegraphics[width=0.4\textwidth]{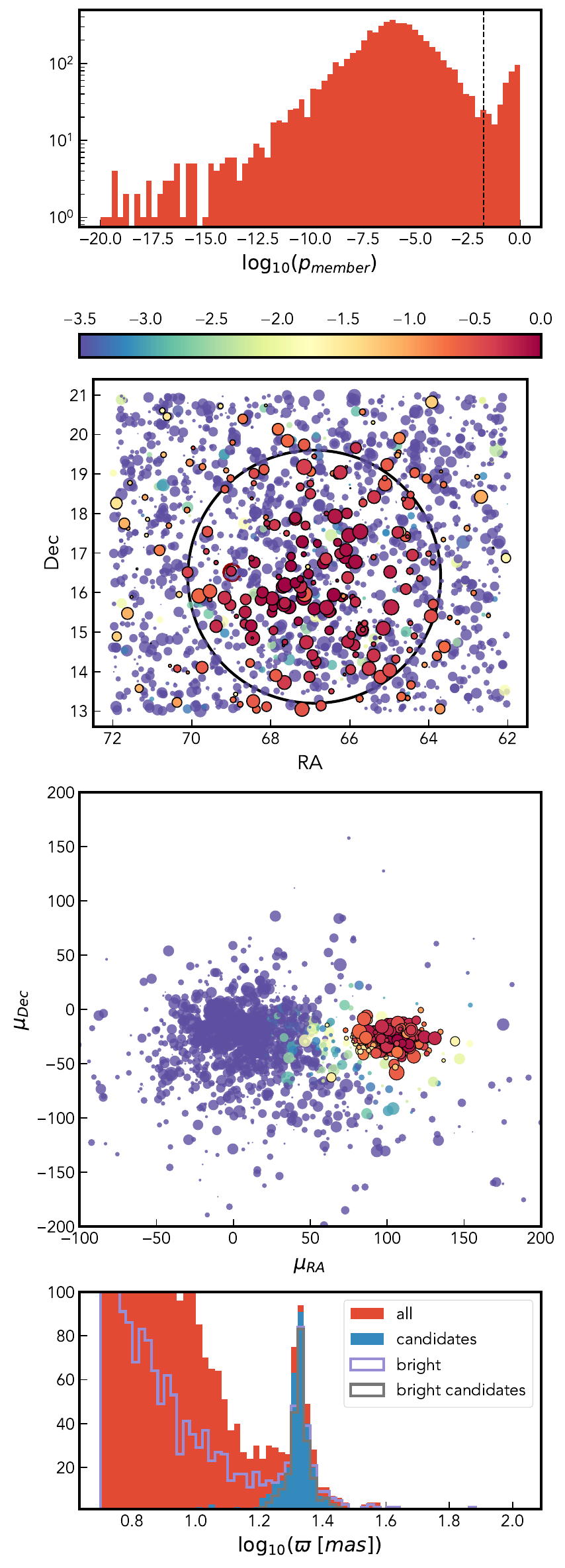}
    \caption{Cluster membership probability for stars in the Hyades field based on eq. \protect\ref{eq:membership_probability}. We show the distribution for all stars (top) and, based on this, the cut at $\log_{10}(p_{\rm{member}})=-1.75$ (vertical dashed line). The middle two panels show the position and proper motion distribution (similar to Fig. \protect\ref{fig:hyades_field}) colored by $\log_{10}(p_{\rm{member}})$. Stars with values greater than -1.75 are shown with black outlines. The bottom panel shows the parallax distribution of all stars in our field and our candidates. 
    }
    \label{fig:hyades_probability}
\end{figure}

\begin{figure*}
\includegraphics[width=0.910\textwidth]{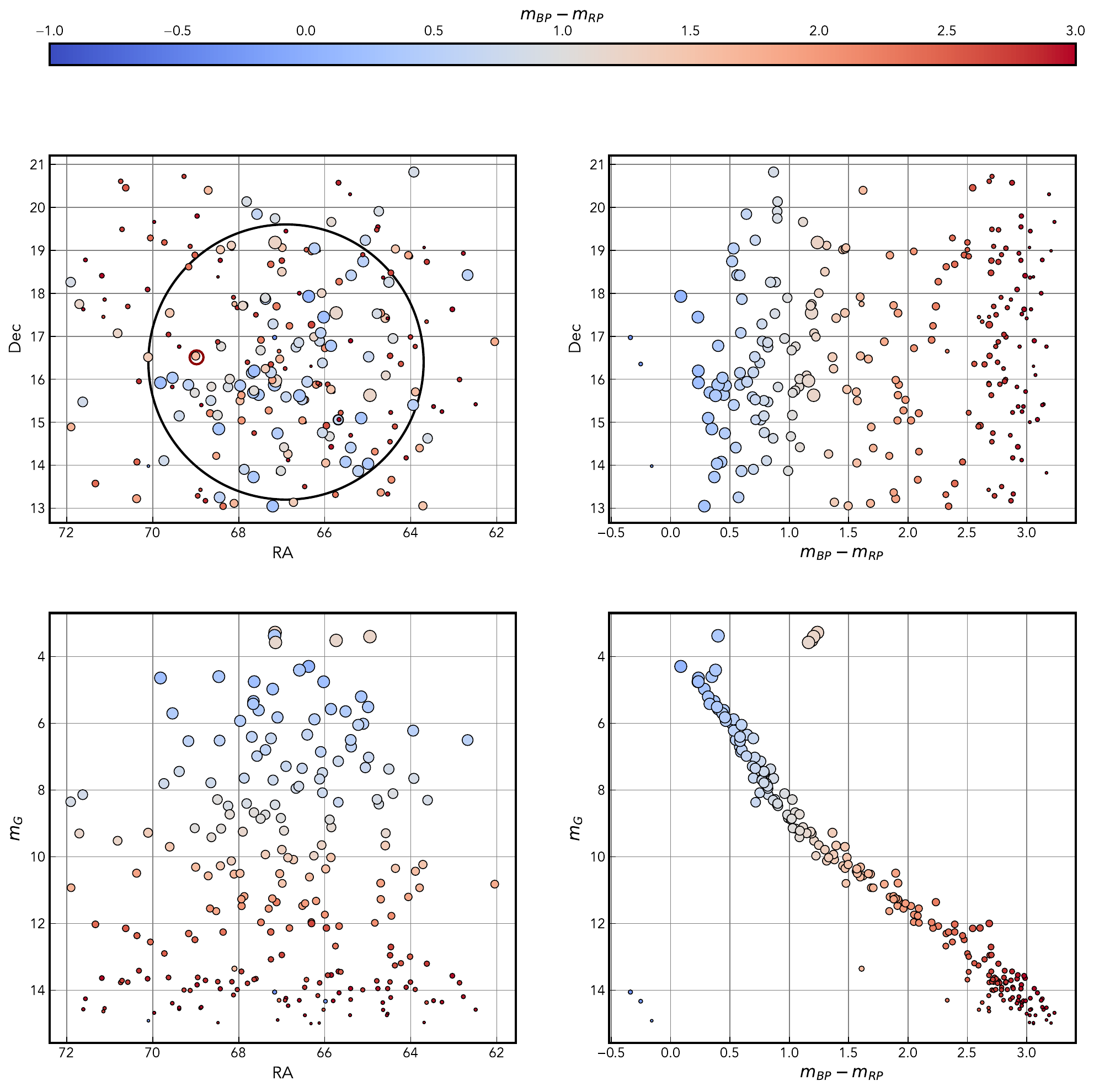}
    \caption{Sky maps and color-magnitude diagrams for the Hyades candidates, colored by {\it Gaia} color. 
    }
    \label{fig:hyades_colour}
\end{figure*}

The value of $p_{\rm member}$ for stars in the field is shown in Fig.~\ref{fig:hyades_probability} from which we choose a critical value of $\log_{10}(p_{\rm member})=-1.75$ giving 229 members which can be seen and identified on the Hertzsprung-Russell diagram shown in Fig. \ref{fig:hyades_colour}.

\begin{table}
\centering
\begin{tabular}{c | c c c c c}
  & $\varpi$ & RA & Dec & $\mu_{RA^*}$ & $\mu_{Dec}$ \\ 
 \hline
 $x_0$ & 22 & 66.9 & 16.4 & 105 & -25 \\
 $\sigma_{x_0}$ & 7 & 3.2 & 3.2 & 35 & 30
\end{tabular}
\caption{Values for $\varpi$, RA, Dec, $\mu_{RA^*}$, $\mu_{Dec}$, and their reported uncertainty in the \textit{Gaia} catalog.
}\label{tab:hyades_params}
\end{table}

\subsubsection{Astrometric and spectroscopic noise}

Following the method introduced in \citet{Andrew22}, we can use the astrometric and spectroscopic noise associated with the measurements in the {\it Gaia} source catalog (which assumes every star is single) to identify and characterize binary systems. This is possible for binaries with periods from days to years, as these can show significant deviations from expected single-body motion. As {\it Gaia} takes many high-precision measurements, the discrepancy between the expected and observed error behavior is predictable and, as we will do here, can be used to estimate periods, mass ratios and companion masses.

The first step is to select systems with significant excess noise. For astrometry, we can use a property directly recorded in the catalog, named $RUWE$. This is equal to the square root of the reduced chi-squared of the astrometric fit and should, for well-behaved observations, give values clustered around 1. Values significantly above 1 suggest that either the model is insufficient, the error is underestimated, or there are one or more significant outlying data points. 
Given that binary systems are ubiquitous (a simple rule-of-thumb is that around half of most samples of sources host more than one star, see for example \citealt{Offner22}), these will be the most common cause of excess error, especially in nearby well-characterized systems outside of very dense fields.

\begin{figure*}
\includegraphics[width=0.910\textwidth]{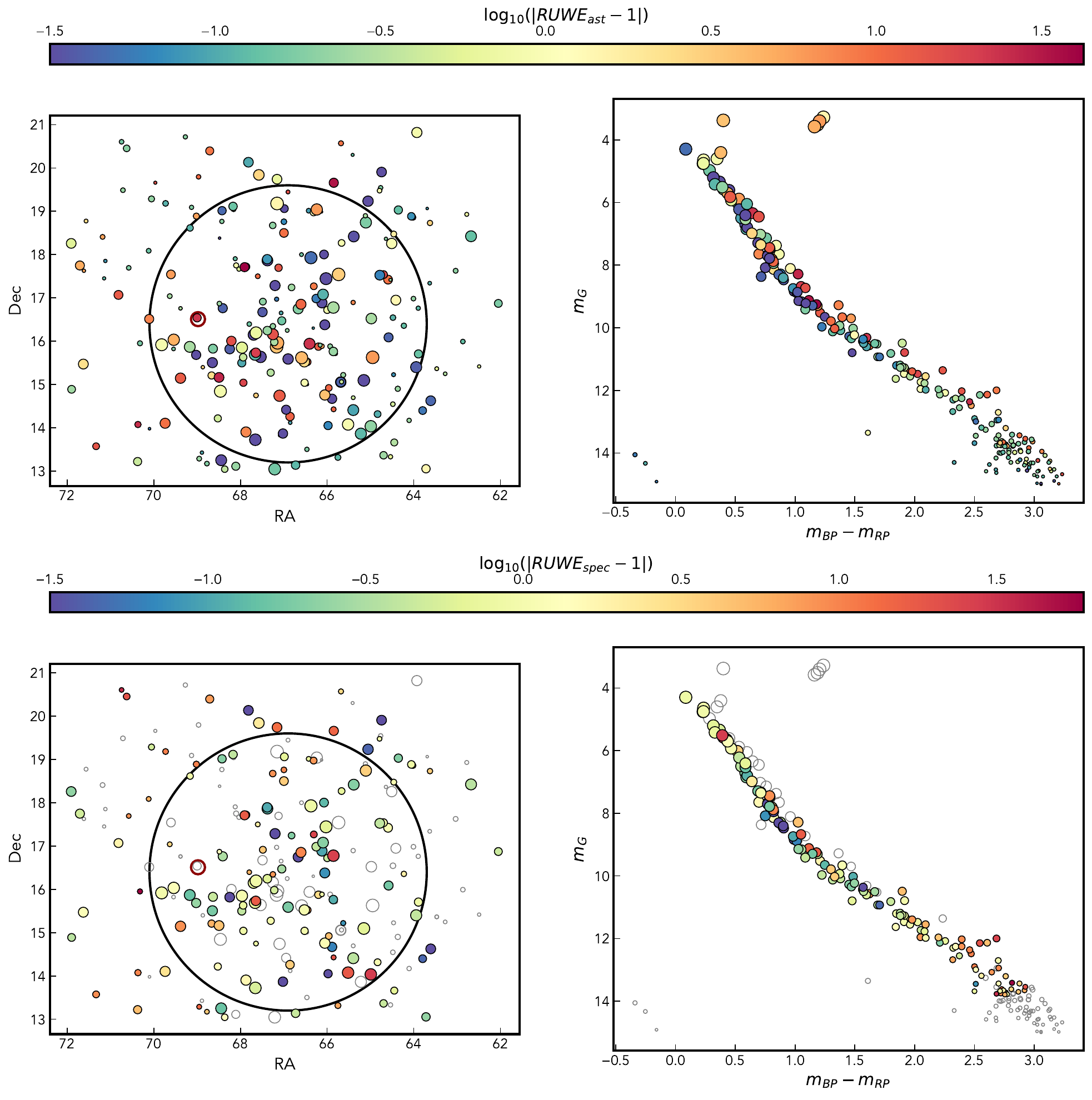}
    \caption{Hyades candidates colored by astrometric (top) and spectroscopic (bottom) renormalized-unit-weight-error ($RUWE$). Values significantly above 1 suggest that the system has an extra source of noise, most ubiquitously a binary companion. Many sources don't have radial velocity measurements in the {\it Gaia} source catalog, and these are denoted with empty grey circles in the bottom plot.
    }
    \label{fig:hyades_ruwes}
\end{figure*}

It is possible to compute a reduced-chi-squared for any quantity where we know the observed variance, expected precision, and the degrees of freedom - and thus we can find the $RUWE$ associated with spectroscopic measurements as well. To do this, we need to estimate the observational measurement error, which we do as a function of the stars' magnitude and color (as detailed in \citealt{Andrew22}) giving $\sigma_{spec}(m_G,m_{BP}-m_{RP})$, the uncertainty expected for a single measurement for each source. Thus we can construct a spectroscopic renormalized unit-weight error, which we'll call $RUWE_{spec}$ to use alongside the astrometric which we'll denote as $RUWE_{ast}$.
These values are shown for Hyades candidate members in Fig. \ref{fig:hyades_ruwes}.

Only a minority of {\it Gaia} sources have radial-velocity observations, which can be missing because sources are too bright ($m_G \lesssim 4$, as seen at the top of the HR diagram), too dim ($m_G \gtrsim 14$, as seen at the bottom), in too dense neighborhoods, or if they are double-lined (with visible absorption lines in more than one of a multiple system, as may be the case with some likely multiple stars above the main-sequence). We use only systems with \texttt{rv\_method\_used = 1} as only these are easily invertible to give binary properties (\citealt{Andrew22} for more details).

The particular value at which $RUWE$ is deemed significantly must be decided pragmatically, and we adopt the values from \citet{Andrew22} of $RUWE_{ast}>1.25$ and $RUWE_{spec}>2$, where the higher criteria for spectroscopic measurements stems from the smaller number of measurements per star and thus the wider spread in $RUWE$. We select sources satisfying both of these criteria as candidate Hyades binaries, giving 56 systems. 
There are some sources that exceed one of these criteria and not the other, and these are interesting potential candidates, but they cannot be used for the next step in the analysis. Using both (generally independent) checks should significantly reduce our number of false positives. It is worth noting that radial-velocity signals are largest for short-period orbits, whereas astrometric signals are largest for systems whose periods match the time baseline of the survey (34 months for {\it Gaia} DR3). This both tells us about which systems we might miss or might meet one criterion and not the other. It also gives the explanation for one of the largest sources of contaminants in this process: triples (or higher multiples) where each significant excess noise comes from a different orbit and thus the two cannot be easily combined or compared.

If we know the $RUWE$ and the measurement error, and assume that all excess noise comes from the contribution of the binary we can invert to find specifically the contribution of the binary:
\begin{equation} \label{eq:sigmabspec}
\sigma_{b,spec} = \sqrt{RUWE_{spec}^2 - 1} \cdot \sigma_{spec}(m_G,m_{BP}-m_{RP}).
\end{equation}
and
\begin{equation} \label{eq:sigmabast}
\sigma_{b,ast} = 2\sqrt{RUWE_{ast}^2 - 1} \cdot \sigma_{ast}(m_G)
\end{equation}
where the factor of 2 comes from the fact that {\it Gaia} takes one-dimensional measurements of the stars 2D position.

\subsubsection{Binary properties from excess error}
The contributions in eqs. \ref{eq:sigmabspec} and \ref{eq:sigmabast} can be mapped back to the properties of the binary and inverted to give the period and (after estimating the mass of the primary) the mass of the companion, as detailed in \citet{Andrew22}. For binary periods less than or equal to the time baseline of the survey the period is approximately:
\begin{equation}
P=\frac{2\pi A}{\varpi}\frac{\sigma_{b,ast}}{\sigma_{b,spec}},
\end{equation}
and the mass ratio follows:
\begin{equation}
q^3 - \alpha q^2 -2\alpha q - \alpha =0,
\end{equation}
where
\begin{equation}
\alpha = \frac{A}{G M \varpi} \sigma_{b,spec}^2 \sigma_{b,ast}
\end{equation}
and $A=1$ AU. $M$ is the mass of the primary star which can be estimated via:
\begin{equation}
M=10^{0.0725(4.76-m_{\mathrm{G}})},
\end{equation}
where $m_{\mathrm{G}}$ is the absolute magnitude of the star \citep{Pittordis19}. This is only strictly relevant for main-sequence stars - but all evolved systems in the Hyades are too bright for \textit{Gaia} spectroscopic measurements and thus will not be included in later analysis (with the exception of white dwarfs, which are too dim).

\begin{figure*}
\includegraphics[width=\textwidth]{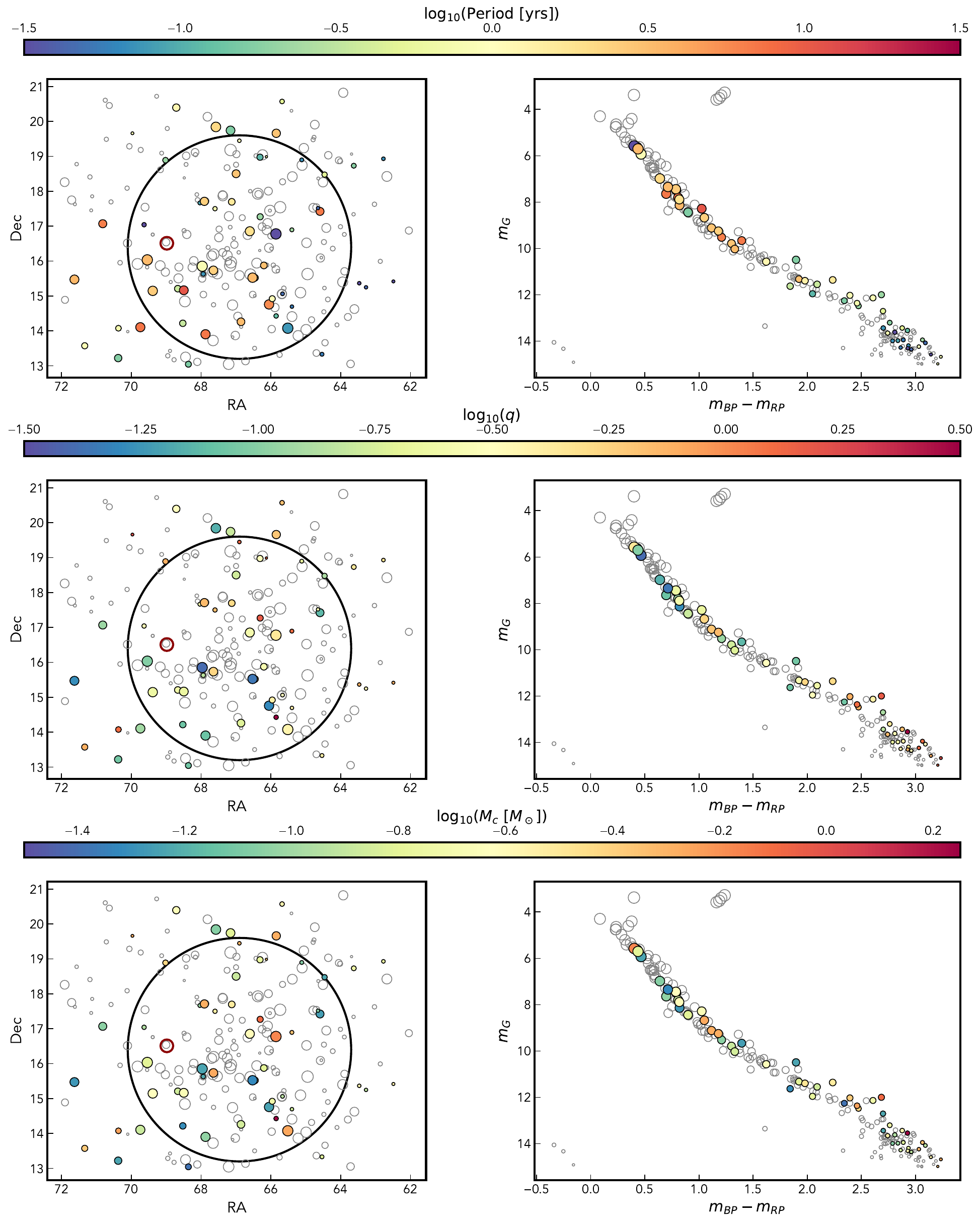}
    \caption{Periods, mass ratios ($q$) and companion masses ($M_\mathrm{c}$) of Hyades candidates inferred from astrometric and spectroscopic $RUWE$. Only sources with significantly high $RUWE$ in both measurements are included here, and all others are shown with empty grey circles.
    }
    \label{fig:hyades_properties}
\end{figure*}

These equations assume the companion has negligible luminosity of its own. If this assumption doesn't hold then the period is slightly overestimated and the mass ratio (and companion mass) are slightly underestimated (see Fig. 3 of \citealt{Andrew22} for more detailed behaviour).
The inferred properties of all 56 systems are shown in Fig. \ref{fig:hyades_properties} and recorded in Tab. \ref{tab:binary_candidates}.

There are some simple consistency checks we can apply to these results. Primarily we know that astrometric measurements should only be discerning for binaries with periods from months to decades \citep{Penoyre22} - thus any deep blue or deep red points are likely spurious solutions - though there are only a handful that have erroneous seeming periods.

As we are searching for significant-mass BHs, we focus on the sources with the highest values of $q$ and $M_\mathrm{c}$, but we should be careful as this is equivalent to selecting those with the largest errors and thus possibly those most likely to truly be erroneous (rather than caused by a binary). For example, the highest mass ratio ($q>1$) sources are amongst the dimmest and thus least reliably measured in the sample - these could be physical, most likely white dwarf companions - but could also be random error.
The brighter stars that show evidence of companions have relatively modest properties - mass ratios below 1 and companion masses significantly below those of a clear BH companion.

Given the period constraints on binaries including BHs present in the simulations, as presented in Fig. \ref{fig:aP}, it is not shocking that we do not find any likely companions. We certainly cannot rule out that these or other stars in the Hyades might have massive compact companions on smaller or wider orbits that {\it Gaia} would be insensitive to. Instead, we are pleased to be able to present a list of candidate binaries whose companions are most likely similar main-sequence stars or white dwarfs.

Stars with massive companions may still be identifiable via their velocity offset. The orbital velocity of a $1.5\,\msun$ star in a binary with a companion of $15\,\msun$ and a period of $10^3(10^4)\,$yr has an orbital velocity of $\sim7(3)\,$km/s. Searching for these systems from velocity offsets is beyond the scope of this work but is an interesting avenue for future exploration.

\subsection{Implications for gravitational waves}

Given the vicinity of the Hyades, it is interesting to ask the question whether a BBH in the Hyades would be observable as a continuous gravitational wave source with ongoing or future experiments. 
Let us therefore adopt a BBH with component masses of $m_1=m_2=10\,\msun$, an average stellar mass of $\langle m\rangle=0.5\,\msun$ and an escape velocity from the centre of the cluster of $v_{\mathrm{esc}}=0.5\,$km/s. Then we assume that the semi-major axis is $a=a_{\rm min}=2~{\rm AU}$, i.e. the minimum before it is ejected in an interaction with a star (eq.~\ref{eq:amin}). This is the most optimistic scenario, because it results in the smallest $a$, but since the interaction time between stars and the BBH goes as $1/a$, a BBH spends a relatively long time at this final, high binding energy. 
An estimate of the absolute duration can be obtained from the required energy generation rate
\citep{2020MNRAS.492.2936A}, from which we find $\sim5\,$Gyr. Because this is much longer than the Hyades' age, it is a reasonable assumption that a putative BBH is near this highest energy state.
For the adopted parameters, $a_{\rm min}\simeq2~$AU. For a typical eccentricity of $\sim0.7$, the peak frequency ($\sim5\times10^{-4}\,$mHz, eq.~37 in \citealt{2003ApJ...598..419W}), i.e. below the lower frequency cut-off of LISA  ($\sim0.1$mHz) and the orbital period of $\sim0.7~$yr is comparable to 
the maximum period that can be found by LISA \citep[$\sim0.7\,$yr,][]{2017ApJ...842L...2C}.
Only for eccentricities $\gtrsim0.99$ (2\% probability for a thermal distribution) the peak frequency is $\gtrsim0.1~$mHz.
BH masses ($\gtrsim30\,\msun$) result in orbital periods comfortably in the regime that LISA could detect ($\lesssim0.08~$yr), but such high masses are extremely unlikely given the high metallicity of the Hyades. 

Because of the low frequency, we consider now whether a BBH in Hyades is observable with the Pulsar Timing Array (PTA). \citet{2005ApJ...627L.125J} show that a BBH at a minimum distance to the sight line to a millisecond pulsar (MSP) of 0.03 pc ($\sim3~$arcmin for the Hyades' distance)  causes a time-of-arrival fluctuation of 0.2-20 ns, potentially observable \citep{2001Natur.412..158V}. Unfortunately, the nearest MSP in projection is PSR J0407+1607 at 5.5 deg\footnote{ATNF Pulsar Catalog by R.N. Manchester et al., at \href{http://www.atnf.csiro.au/research/pulsar/psrcat}{http://www.atnf.csiro.au/research/pulsar/psrcat}}. If the BBH was recently ejected, it may be close to a MSP in projection, but the maximum distance a BBH could have travelled is $\sim1~{\rm deg}$ (Section~\ref{sec:radial_distribution}) and there are only 4 pulsars within a distance of 10 deg, so this is unlikely as well.
In conclusion, it is unlikely that (continuous) gravitational waves from a BBH in or near the Hyades will be found. 


\subsection{Gravitational microlensing}
Because of the vicinity of the Hyades, BHs have  relatively large Einstein angles and we may detect a BH or a BBH through microlensing. For a BH mass of $10\,\msun$ at a distance of 45 pc and a source at 5\,kpc, the Einstein angle is $\theta_{\rm E}\simeq40\,$mas. Assuming that background stars in the galaxy are distant enough to act as a source, we find from the {\it Gaia} catalogue that the on-sky density of background sources is $\Sigma_{\rm S}\simeq10^{-9}\,{\rm mas}^{-2}$. The Hyades moves with an on-sky velocity of $v_{\rm H}\simeq100\,{\rm mas\,yr}^{-1}$ relative to the field stars. This gives us a rough estimate of the microlensing rate of $R\simeq 2\theta_{\rm E}N_{\rm BH}\Sigma_{\rm S}v_{\rm H}\simeq2\times10^{-5}\,{\rm yr}^{-1}$, where we used $N_{\rm BH}=2$.
Even if we consider astrometric lensing, for which the cross section for a  measureable effect is larger  \citep[for example][]{Paczynski1996, 1996ApJ...470L.113M}, the expected rate is too low. This is mainly because of the low number of background sources because of Hyades' location in the direction of the Galactic anticentre. 
Perhaps the orders of magnitude higher number of stars that will be found by  LSST can improve this. More promising in the short term is to search for BHs in other OCs which are projected  towards the Galactic centre. 



\section{Conclusions} \label{sec:conclusions}
In this study, we present a  first attempt to find dynamical imprints of stellar-mass black holes (BHs) in Milky Way open clusters. In particular, we focused on the closest open cluster to the Sun, the Hyades cluster. We compared the mass density profiles from a suite of direct $N-$body models, conceived with the precise intent to model the present-day state of Hyades-like clusters (\citealp{wang2021}), to radial mass distributions of stars with different masses, derived from \textit{Gaia} data (\citealp{evans2022}). 

Our comparison favors $N-$body models with $2-3$ BHs at present. In these models, the presence of a central BH component quenches the segregation of visible stars, and leads to less concentrated distributions. Star clusters with $2-3$ BHs (and a BH mass fraction $f_{\mathrm{BH}}\simeq0.1$) best reproduce the observed half-mass radius, while those that never possessed BHs display a value that is $\sim 30\%$ smaller. This result is further confirmed by the radial distribution of high-mass stars ($m\geq 0.56 \, \msun$), which, being more segregated, are more affected by the presence of central BHs. Models in which the last BH was ejected recently ($\leq 150$ Myr ago) can still reproduce the density profile. For these model, we estimate that the ejected (binary) BHs are at a typical distance of $\sim 60$ pc from the Hyades. 

Models with $2-3$ BHs have a one-dimensional dispersion in the innermost parsec of $\sim350\,$m/s compared to  $\sim250\,$m/s for the no BH case and both are consistent with the available data. The tidal tails of models with and without BHs are almost indistinguishable.

In absence of primordial binaries, about 94\% of the BHs in the present-day state of our $N-$body models dynamically couple with other objects and form binary and triple systems. Among them, 50\% of the clusters with BHs host BH-star binary systems. Their period distribution peaks at $\sim 10^3$ yr making it unlikely to find BHs through velocity variations. We explored the possible candidate stars with a BH companion, based on their excess error in the \textit{Gaia} singe-source catalog but otherwise high membership probability. We found 56 possible binaries candidates, but none which show strong evidence of sufficient companion mass to be a likely BH. 
Also, we explored the possibility to detect binary BHs through gravitational waves with Pulsar Timing Array. We found that (continuous) gravitational waves from a BBH in or near the Hyades is unlikely to be found. Finally, we estimated that detecting dormant BHs with gravitational microlensing is unlikely too.

Our study suggests that, at the present day, the radial mass distribution of stars provides the most promising discriminator to find signatures of BHs in open clusters. In particular, the most massive stars within the cluster, and their degree of mass segregation, represent the best tracers for the presence of central BHs. For the case of the Hyades, its present-day structure requires a significant fraction of BHs to form with kicks that are low enough to be retained by the host cluster.

Our approach of detailed modelling of individual OCs can be applied to other OCs to see whether Hyades is an unique cluster, or that BHs in OCs are common. Charting the demographics in OCs in future studies will be a powerful way to put stringent constraints on BH kicks and the contribution of OCs to gravitational wave detections. 

\section*{Acknowledgements}
We thank the anonymous referee for the useful comments, which helped to improve the quality of this manuscript.
ST acknowledges financial support from the European Research Council for the ERC Consolidator grant DEMOBLACK, under contract no. 770017.
MG acknowledges support from the Ministry of Science and Innovation (EUR2020-112157, PID2021-125485NB-C22, CEX2019-000918-M funded by MCIN/AEI/10.13039/501100011033) and from AGAUR (SGR-2021-01069). ZP acknowledges that this project has received funding from the European Research Council (ERC) under the European Union’s Horizon 2020 research and innovation programme (Grant agreement No. 101002511 - VEGA
P). 
LW thanks the support from the one-hundred-talent project of Sun Yat-sen University, the Fundamental Research Funds for the Central Universities, Sun Yat-sen University  (22hytd09) and the National Natural Science Foundation of China through grant 12073090 and 12233013.
FA acknowledges financial support from MCIN/AEI/10.13039/501100011033 through grants IJC2019-04862-I and RYC2021-031638-I (the latter co-funded by European Union NextGenerationEU/PRTR).
ST thanks Michela Mapelli for valuable comments and suggestions.

\section*{Data Availability}
The data underlying this article will be shared on reasonable request to the corresponding authors.



\bibliographystyle{mnras}
\bibliography{references} 
\bsp	
\label{lastpage}

\begin{landscape}
\begin{table}
\scriptsize\begin{filecontents*}{barcelona_hyades_data.csv}
sourceid,RA,Dec,parallax,pmra,pmdec,mg,bprp,RUWEast,sigmabast,RUWEspec,sigmabspec,$P$,$q$,$M$c
3313842529024545664,65.85493019594335,16.777147465731876,20.795046,106.97092,-26.247469,5.573977,0.40038252,1.2823561,0.41557395,27.358295,23.794285,0.025030272,0.44117424,0.6805129
3311000119668435200,65.51518023603637,14.077103275227195,22.026937,113.774475,-21.321016,5.6442456,0.42680836,1.8872856,0.7889371,19.065733,14.198232,0.075180255,0.3714967,0.5546593
3309990523179108480,69.53972879902068,16.03321720295417,17.820915,84.96662,-8.793861,5.7026787,0.4365778,5.6922646,2.4559207,2.1845562,1.2523471,3.2795131,0.09720816,0.15520167
3312637842237882880,67.96612153925633,15.851471350184573,21.005054,99.49563,-34.32999,5.928877,0.46518755,1.5450443,0.47656766,2.0089986,0.8186548,0.82594234,0.039998896,0.057937805
144130516816579200,67.57532736919829,19.84041406012967,19.89243,86.11872,-26.106121,6.9827037,0.6383934,3.7624655,1.3966898,3.2055814,0.9599861,2.179701,0.069570266,0.08619921
3312751882209460864,66.52490368958763,15.524230392191061,23.521027,124.196014,-25.313429,7.3494782,0.7129674,3.4419775,1.343747,2.0151875,0.5174881,3.2901115,0.044089917,0.04835587
3309493720019304576,69.38373895275495,15.146378860785259,25.64366,101.49229,-18.512648,7.4436464,0.7864218,13.72717,5.776902,9.669819,2.558703,2.6238859,0.2289321,0.23954499
3311151233797884416,66.05246947657736,14.758157583423541,20.856197,110.226265,-12.401878,7.4817166,0.7890749,4.709577,1.9607697,2.121658,0.4972577,5.634556,0.05053243,0.056626927
3307815006281475200,67.87268564180225,13.903399057854472,20.276747,86.607285,-6.108541,7.643193,0.69734,8.31893,3.525965,2.614277,0.7415865,6.9882374,0.083096355,0.09157219
3309170875916905856,69.73925748852255,14.105477042258816,18.390837,91.53597,-20.422842,7.8071556,0.80762196,10.741918,4.6240096,3.6913257,1.0178571,7.361735,0.118229695,0.1313378
3313743710417140864,66.60303643900508,16.853180488088583,26.757921,95.673996,-38.761845,7.8861456,0.81799173,12.058122,5.1952686,10.748748,3.0819666,1.8774892,0.25809193,0.24698886
3309006602007842048,71.62700947833584,15.471931785895292,20.5006,87.325676,-25.352728,8.138844,0.8205805,3.2997699,1.3771452,2.1922784,0.5878727,3.40549,0.052342582,0.052890003
3309540170088830464,68.49438499286096,15.16352719505778,20.641268,108.03673,-33.110138,8.2856655,1.0252385,27.214418,11.50729,5.577302,1.5087116,11.012355,0.22136614,0.2177251
48061409893621248,67.1555527872302,19.74051990049017,20.810343,102.9064,-38.101357,8.441752,0.89888763,1.6440603,0.53488284,14.239208,4.5749774,0.16743268,0.16246925,0.15522723
3312631623125272448,67.64578394363194,15.733880135447295,17.485168,100.58648,-25.256905,8.67361,1.0493889,21.024082,7.626148,18.202843,5.276344,2.4634974,0.55225676,0.54067534
48197783694869760,65.84566930782594,19.658531350570186,21.598646,97.3178,-37.780178,9.1158905,1.1155329,34.355576,10.966481,16.315971,5.2403026,2.887581,0.62754923,0.52858657
3314151251273992832,67.9051025367819,17.70963694563987,28.507998,98.723816,-22.038061,9.247976,1.1795635,43.27126,13.478946,17.999285,5.767014,2.443363,0.70073265,0.5220939
3406216383523959936,70.81584857548913,17.068982088574,22.40773,84.4349,-41.66873,9.519791,1.2086744,14.337425,4.4526625,2.7862427,0.90990114,6.508449,0.11398687,0.08856268
47319858019739264,64.59216618931373,17.421415275740294,23.752161,104.46378,-50.567753,9.66315,1.3945608,10.326841,3.221443,2.0533266,0.5846881,6.9130993,0.07403989,0.054991435
3314484094059409408,66.99617799986078,18.500100518411095,19.760468,108.59516,-40.885216,9.7873955,1.2999792,10.631993,3.3175225,4.8615503,1.7125669,2.9215875,0.1696449,0.13192531
3310876802567157248,66.85605547720428,14.260578556388577,20.42228,102.45836,-19.95602,10.026131,1.33076,12.669773,3.9585905,4.7959375,1.7974386,3.2138977,0.18906817,0.13960828
3308127405023027328,70.3740966937844,13.221124178946916,22.662909,99.23287,-18.63009,10.491406,1.8964214,1.4050351,0.29435474,5.7457695,2.2721488,0.1703603,0.08793274,0.057852447
3410898202818634624,68.70917414686971,20.394419352148518,18.76046,110.18581,-22.617853,10.571909,1.6212931,9.876004,2.8892152,8.526632,3.900182,1.1767963,0.32126498,0.2233311
3313526556869575296,66.20070596676636,15.874649203665436,20.678333,100.34452,-12.269188,11.324224,1.9229269,12.951795,4.8260064,3.9638534,2.295628,3.0298467,0.2657315,0.15727599
3314185129975960448,67.12038862642493,17.695791947186837,20.918434,109.48695,-44.22729,11.360385,2.2351828,15.092351,5.5574145,8.521062,4.8796754,1.6225665,0.52000636,0.3046414
3312751229374292864,66.45571421241527,15.521152733202104,19.244942,118.904,-10.06363,11.395764,1.9788427,13.038091,4.7377243,8.969254,5.425513,1.3522661,0.54579365,0.32762027
3309541170817293824,68.66689038204488,15.208964911975178,21.075073,103.15689,-35.85557,11.549942,2.0912971,3.344131,1.0991819,6.668379,4.2592897,0.36493292,0.2443392,0.13831078
3309336867813205760,68.52256353868661,14.21747714186208,18.199448,85.34915,-16.904617,11.629388,1.8422108,1.4415438,0.34694323,2.4431305,1.6714453,0.3399052,0.08429948,0.049661882
47917816253918720,66.31113667576857,18.973455972926324,18.569708,97.28108,-28.887608,11.957183,2.0487833,2.0377967,0.5246777,8.945428,7.500937,0.11225916,0.30151984,0.16694689
3313958046465282304,66.30695338557209,17.268076545714347,19.312836,107.2612,-27.188425,11.998245,2.6850185,9.333254,2.704557,26.169952,19.287603,0.2163822,1.5024871,0.81455195
3308152208457011072,71.33211010053422,13.574216271157622,19.19447,87.72377,-9.913272,12.024581,2.3941412,15.202123,4.3817663,9.862589,7.8390193,0.86788297,0.77107835,0.41712388
3311162843092490496,65.96041133865042,14.921402707495972,25.659887,113.94731,-28.725637,12.134732,2.6072578,11.986289,3.3206553,5.654087,4.5110784,0.85494554,0.39180115,0.18730214
3307489031146701696,68.36303444839065,13.045385968619557,23.333511,110.45561,-21.440964,12.256755,2.342475,1.363789,0.2545818,2.3844779,2.0315888,0.16005203,0.08510464,0.041261565
3308428877366857728,70.36635522147463,14.076016738839469,26.182947,99.25034,-33.95699,12.363147,2.4602451,25.720503,7.026489,10.856859,10.731864,0.745237,1.2423422,0.56752026
3410409886514680064,69.01782417867581,18.888423253705223,20.245579,84.23755,-22.036737,12.484824,2.47365,6.511168,1.7505827,11.434649,12.022853,0.21433577,0.76413935,0.3754725
47729215646206080,64.46150286122511,18.475053544960033,21.640215,113.170876,-30.777138,12.703422,2.6998405,2.711683,0.7202027,2.0415013,2.0097606,0.49351305,0.12772296,0.059066143
48455894049191424,63.624805023608,18.729881346946026,20.680376,110.10128,-30.568058,13.204377,2.7620544,3.769246,0.98915327,4.831599,7.9032397,0.18036395,0.43028054,0.18605632
49157794785448192,65.67911798706223,20.56998136359179,22.204473,109.14224,-37.47773,13.435896,2.8771925,10.13076,2.8767006,4.160689,7.6917205,0.50197315,0.66635245,0.2701574
3312613000147191680,67.93658293586253,15.630149907181586,21.92811,104.99937,-31.718637,13.436778,2.7022161,1.2680652,0.22256374,2.1944704,4.0981297,0.073810354,0.14576809,0.059358526
3311038666998173184,65.84974595899124,14.427849817287893,21.160046,110.94352,-21.135527,13.543492,2.9259453,12.498561,3.6824527,20.370281,41.5711,0.12476101,5.761253,2.334608
3313207698500079360,69.63189917287164,17.04055509824066,22.853317,106.24673,-31.3891,13.618415,2.808816,1.2659032,0.23495989,4.2582116,9.578849,0.031987548,0.282138,0.109801285
3314140599755959424,67.60020695729807,17.499587834144798,18.23571,78.77907,-28.804348,13.654834,2.7426186,10.872717,3.3146904,2.9538498,6.8027906,0.79631144,0.6842594,0.28725055
3313839574086963712,65.39849909000245,16.894192274532255,24.33773,125.71914,-40.65925,13.948855,3.064352,8.383781,2.817,5.9499407,14.526592,0.23746127,1.244193,0.44788775
3310478710637066752,64.53629289150157,13.331259196596612,20.93439,111.15798,-14.546099,13.949699,2.8872652,1.5854778,0.4164988,4.107172,10.453021,0.056723293,0.3978708,0.15124361
3314137846679343872,68.0336649745617,17.6643440638911,21.075275,99.88685,-30.655151,13.979582,2.8372974,1.5255272,0.3941187,2.4364855,6.2334156,0.08940822,0.25801963,0.097356565
47760521664503296,65.11560537190881,18.89705576636259,19.672964,104.719475,-32.569706,13.988257,2.7881794,1.2785263,0.27335688,2.2563696,5.902441,0.070158176,0.21893603,0.08457476
3311221258944195200,65.66660974089085,15.062399584369444,23.376665,84.40576,-23.434784,14.077939,3.0911245,1.3902847,0.34179467,3.8369422,9.311588,0.046796,0.32785812,0.11720649
48509186003340416,62.776059616737115,18.928963628615932,19.93843,109.657745,-28.472279,14.192208,2.9350224,1.7115903,0.5110402,3.5404887,11.017797,0.069329664,0.46646634,0.17331943
48048902948906368,66.90245867179468,19.445767441293732,19.231146,96.80087,-38.773808,14.238761,3.0310307,16.237202,6.0698442,3.4705567,9.437445,0.99670756,1.3147694,0.4911199
3311102820926310784,65.39669553192954,14.695153868168017,20.326693,106.496574,-19.731718,14.278773,2.862668,1.3977338,0.3713231,2.7254941,9.875319,0.05512948,0.37344685,0.1358137
3311810047420637568,63.46857579356195,15.36487063340007,22.387419,123.95292,-13.587856,14.308949,2.9282503,1.5574532,0.4591437,5.319673,20.914965,0.029223839,0.7675065,0.26816785
3311804515502788352,63.273952149488956,15.247700118856342,22.357246,122.44951,-18.603453,14.355454,2.9613495,1.3029861,0.32675368,2.5205526,10.018484,0.04347603,0.35185495,0.12204755
47322297561027712,64.65025269592638,17.51585330705525,27.013884,64.83592,-61.09387,14.369595,2.959774,1.2812126,0.31491673,2.0114625,7.8232236,0.04440918,0.27206907,0.08790898
45198178534988672,62.49039974488908,15.416883616520483,21.22679,118.5227,-21.854282,14.58098,3.1423988,1.5162789,0.48248684,4.441856,20.385468,0.03322999,0.7961063,0.27099073
3410559832411244160,69.96546164739077,19.6593898461387,23.499067,87.291985,-34.15228,14.677451,3.234291,16.270372,7.1377497,3.6114807,11.025825,0.8210118,1.6573471,0.5350476
47959705067457152,66.12800628155945,18.986786982710726,20.498665,111.01914,-43.00263,14.982163,3.2060423,10.618676,5.2158256,2.239869,10.304867,0.7358764,1.3935447,0.44927633
\end{filecontents*}
\csvautotabular{barcelona_hyades_data.csv}
\caption{Candidate binaries in the Hyades with significant astrometric and spectroscopic excess noise. RA and Dec are in degrees - sigmabspec is in kms$^{-1}$ - all other quantities are expressed in the appropriate combinations of mas, yr and $M_\odot$.}
\label{tab:binary_candidates}
\end{table}
\end{landscape}

\end{document}